\documentclass[preprint,aps,prd,superscriptaddress,nofootinbib]{revtex4-1}

\usepackage{amsmath}
\usepackage{graphicx}
\usepackage{hyperref}
\usepackage[T1]{fontenc}
\usepackage{slashed}
\usepackage{bbold}
\usepackage{xcolor}
\usepackage{url}
\usepackage{placeins}
\usepackage{float}

\newcommand{\Lagr}{\mathcal{L}}
\newcommand{\be}{\begin{equation}}
\newcommand{\ee}{\end{equation}}

\DeclareMathOperator{\Tr}{Tr}

\usepackage[normalem]{ulem}

\definecolor{purple}{rgb}{0.43, 0.31, 0.4}

\begin{document}

\title{Renormalizable Models of Flavor-Specific Scalars}

\preprint{PITT-PACC-2114}

\author{Brian Batell}
\email{batell@pitt.edu}
\affiliation{Pittsburgh Particle Physics, Astrophysics, and Cosmology Center,\\Department of Physics and Astronomy, University of Pittsburgh, Pittsburgh, USA}
\author{Ayres Freitas}
\email{afreitas@pitt.edu}
\affiliation{Pittsburgh Particle Physics, Astrophysics, and Cosmology Center,\\Department of Physics and Astronomy, University of Pittsburgh, Pittsburgh, USA}
\author{Ahmed Ismail}
\email{aismail3@okstate.edu}
\affiliation{Department of Physics, Oklahoma State University, Stillwater, OK 74078, USA}
\author{David McKeen}
\email{mckeen@triumf.ca}
\affiliation{TRIUMF, 4004 Wesbrook Mall, Vancouver, BC V6T 2A3, Canada}
\author{Mudit Rai}
\email{mur4@pitt.edu}
\affiliation{Pittsburgh Particle Physics, Astrophysics, and Cosmology Center,\\Department of Physics and Astronomy, University of Pittsburgh, Pittsburgh, USA}

\begin{abstract}

New light singlet scalars with flavor-specific couplings represent a phenomenologically distinctive and flavor-safe alternative to the well-studied possibility of Higgs-portal scalars. However, in contrast to the Higgs portal, flavor-specific couplings require an ultraviolet completion involving new heavy states charged under the Standard Model gauge symmetries, leading to a host of additional novel phenomena. Focusing for concreteness on a scenario with up quark--specifc couplings, we investigate two simple renormalizable completions, one with an additional vector-like quark and another featuring an extra scalar doublet. We consider the implications of naturalness, flavor- and CP-violation, electroweak precision observables, and direct searches for the new states at the LHC.  These bounds, while being model-dependent, are shown to probe interesting regions in the parameter space of the scalar mass and its low-energy effective coupling, complementing the essential phenomenology of the low-energy effective theory at a variety of low and medium energy experiments. 

\end{abstract}

\maketitle

%%%%%%%%%%%%%%%%%%%%%%%%%%%%%%%%%%%%%%%%
%%%%%%%%%%%%%%%%%%%%%%%%%%%%%%%%%%%%%%%%
\section{Introduction}

Light dark sectors that couple weakly to the Standard Model (SM) may address some of the key open questions in particle physics today~\cite{Alexander:2016aln,Battaglieri:2017aum,Beacham:2019nyx}. For instance, dark matter may reside in a dark sector, possibly along with other states that are SM gauge singlets, and communicate with the SM through a light mediator particle. One commonly investigated model employs a singlet scalar as the mediator interacting through the Higgs portal~\cite{Silveira:1985rk,McDonald:1993ex,Burgess:2000yq,Krnjaic:2015mbs}. In this scenario, the singlet scalar inherits its interactions with SM matter via mixing with the Higgs boson, thereby coupling preferentially to the heavy third generation fermions and massive electroweak bosons. This leads to a characteristic phenomenology for a light scalar mediator with masses in the MeV-GeV range, with the best probes typically coming from penguin-induced rare meson decays and exotic Higgs decays; see, for example, Ref.~\cite{Krnjaic:2015mbs}.

While the Higgs portal provides a well-motivated and popular benchmark, it is of interest to explore other models with qualitatively distinct patterns of mediator couplings to the SM. Such investigations are warranted by the prospect of novel phenomena and new experimental opportunities to probe dark sectors. For scalar mediators in particular, an immediate obstacle is the specter of new dangerous flavor changing neutral currents (FCNCs). Unlike the Higgs portal, which automatically respects Minimal Flavor Violation~\cite{DAmbrosio:2002vsn}, there is no built-in protection mechanism against large FCNCs for general scalar mediators. From a bottom-up perspective, one can circumvent this issue by appealing to a flavor hypothesis on the structure of the scalar mediator couplings, devised so as to suppress FCNCs at tree level. In this regard, scalar mediators respecting the {\it flavor-specific} hypothesis provide an interesting alternative to the Higgs portal~\cite{Batell:2017kty} (for related work, see Ref.~\cite{Egana-Ugrinovic:2018znw,Egana-Ugrinovic:2019dqu}). Under this hypothesis, the scalar couples to one (or a few) SM fermion mass eigenstate(s) in the physical basis. Particularly if the singlet couples preferentially to first or second generation states, this scenario leads to a distinctive phenomenology compared to the Higgs portal model. This point has been illustrated in previous studies of light hadrophilic dark matter based on an up-specific scalar mediator \cite{Batell:2018fqo} and a possible explanation of the muon anomalous magnetic moment discrepancy~\cite{Muong-2:2006rrc,Muong-2:2021ojo} based on a muon-specifc scalar mediator~\cite{Batell:2017kty}. 

Open questions in this framework related to the short distance structure of the theory remain. Unlike the renormalizable Higgs portal, the flavor specific hypothesis is necessarily formulated in an effective field theory (EFT) setting, where the coupling of interest emerges from a dimension-five operator. Particularly for sizable scalar mediator couplings to matter, we anticipate the presence of new SM-charged degrees of freedom near the weak scale. It is therefore important to study concrete renormalizable completions of flavor-specific EFTs as they can point to additional constraints and experimental prospects associated with the new heavy states. 

In this work we study renormalizable completions of flavor-specific EFTs, focusing for concreteness on models realizing up-quark specific couplings. We study two simple completions of this model, one involving a vector-like quark (VLQ) and another involving a second scalar doublet in addition to the Higgs. We consider the implications of naturalness on the couplings of the light scalar mediator and constraints on the models from  electroweak precision observables, flavor- and CP-violation, CKM unitarity, and searches for new particles at the LHC. We demonstrate that these additional tests, while being model-dependent, can probe new regions of the low energy EFT scalar mass--coupling parameter space. This study therefore builds on and is highly complementary to the previous flavor specific-EFT studies of Refs.~\cite{Batell:2017kty,Batell:2018fqo}. 

Another important open structural question pertains to the ultraviolet dynamics generating the flavor-specific coupling structure. In all likelihood, the resolution of this issue must be tied to the origin of SM flavor, itself a challenging open question. We do not address this issue in this work, but instead focus on the more tractable problem of realizing the flavor-specific EFT in simple renormalizable models and studying their phenomenology.

This paper is organized as follows. In Sec.~\ref{sec:EFT} we review the EFT of the flavor-specific scalar mediator. In Sec.~\ref{sec:VLQ-completion} we study a renormalizable completion with a VLQ, while in Sec.~\ref{sec:SD-completion} we consider one involving a second scalar electroweak doublet. Our conclusions are presented in Sec.~\ref{sec:conclusions}. Appendix \ref{app:hypothesis} describes the flavor hypothesis for each renormalizable completion, while Appendix \ref{app:VLQ-scalar-couplings} provides details of the physical interactions in the VLQ completion. 

%%%%%%%%%%%%%%%%%%%%%%%%%%%%%%%%%%%%%%%%
%%%%%%%%%%%%%%%%%%%%%%%%%%%%%%%%%%%%%%%%
\section{Effective field theory of flavor-specific scalar}
\label{sec:EFT}

In this section, we  review the EFT framework describing a new light scalar $S$ with {\it flavor-specific} couplings,  meaning that the scalar predominantly couples to a particular SM fermion mass eigenstate \citep{Batell:2017kty}. To understand  the flavor-specific hypothesis, it is useful to start from the Yukawa interactions in the SM quark sector:
\be
\label{eq:smlag}
\mathcal{L}_\mathrm{SM} = i \overline{Q}_L \slashed{D} Q_L + i \overline{u}_R \slashed{D} u_R + i \overline{d}_R \slashed{D} d_R - \left( \overline{Q}_L Y_u u_R H_c + \overline{Q}_L Y_d d_R H + \mathrm{h.c.} \right),
\ee
where $Q_L^\top = (u_L, d_L)$ 
and $H$ is the Higgs doublet with $H_c=i\sigma^2H^\ast$. 
The Yukawa interactions in~(\ref{eq:smlag}) break the large $U(3)_Q \times U(3)_U \times U(3)_D$ global flavor symmetry down to baryon number $U(1)_{B}$. In many extensions of the SM there are new couplings that also break the flavor symmetry, leading to the  dangerous prospect of new large FCNCs. It is common to invoke a flavor hypothesis that restricts the form of these new couplings in such a way that new FCNCs are adequately suppressed.  The most common choice is MFV~\cite{DAmbrosio:2002vsn}, which states that the Yukawa couplings $Y_u, Y_d$ are the only flavor-breaking spurions present in the theory, such that all new couplings that break flavor are constructed out of  $Y_u$ and $Y_d$.

The flavor-specific hypothesis takes a different route from MFV to ensure the suppression of new FCNCs. To build up to the flavor-specific hypothesis, one can first understand how the quark flavor symmetry is broken if only one of the Yukawas (up or down) are nonvanishing. In the case of $Y_u \neq 0$ and $Y_d \rightarrow 0$, the $U(3)_D$ symmetry is unbroken, while a general $Y_u$ results in the breaking pattern
\begin{equation}
\label{eq:Yu-break}
U(3)_Q \times U(3)_U \rightarrow U(1)_u \times U(1)_c \times U(1)_t    ~~~~~(Y_u \neq 0, Y_d = 0).
\end{equation}
Similarly, in the case $Y_u \rightarrow 0$ and $Y_d \neq 0$, the $U(3)_U$ symmetry is respected, while general $Y_d$ breaks the symmetry according to
\begin{equation}
\label{eq:Yd-break}
U(3)_Q \times U(3)_D \rightarrow U(1)_d \times U(1)_s \times U(1)_b   ~~~~~(Y_u = 0, Y_d \neq 0),
\end{equation}
In the case of the SM, both $Y_u$ and $Y_d$ are non-vanishing and the CKM matrix is nontrivial. Hence the separate $U(1)^3$ quark flavor symmetries preserved by $Y_u$ (in Eq.~(\ref{eq:Yu-break})) and $Y_d$ (in Eq.~(\ref{eq:Yd-break})) are different, and only the full $U(1)_B$ baryon number symmetry remains.

With this understanding, we now consider an EFT containing a real SM singlet scalar $S$ that dominantly interacts with the SM through a dimension-five operator contained in the Lagrangian
\be
\label{eq:bsmlag}
\begin{aligned}
\mathcal{L}_S = \frac{1}{2} \partial_\mu S \partial^\mu S - \frac{1}{2} m_S^2 S^2 &- \biggl( \frac{c_S}{M} S \overline{Q}_L u_R H_c + \mathrm{h.c.} \biggr).
\end{aligned}
\ee
Under the flavor-specific hypothesis, the coupling $c_S$ only involves a single up-type quark in the mass basis. As an interesting example which we will study throughout this paper, consider the case of an up-specific hypothesis, so that $c_S \propto {\rm diag}(1,0,0)$ in the mass basis. The $U(3)^3$ flavor symmetry is then broken by $c_S$ according to the pattern
\begin{equation}
\label{eq:cS-break}
U(3)_Q \times U(3)_U \rightarrow U(1)_u \times U(2)_{ctL} \times U(2)_{ctR}.
\end{equation}
In particular, simultaneous diagonalization of $c_S$ and $Y_u$ implies that the $U(1)_u$ factor in Eq.~\eqref{eq:cS-break} is the same as the one left unbroken by $Y_u$ in Eq.~\eqref{eq:Yu-break}. We note that the flavor-specific hypothesis can be viewed as a special case of alignment.

The EFT framework provides a good starting point for phenomenological investigations of light flavor-specific scalars, as illustrated by the studies of Ref.~\cite{Batell:2017kty,Batell:2018fqo}. However,  two basic open questions related to the UV structure of the theory remain. First, Eq.~(\ref{eq:bsmlag}) should emerge from a renormalizable theory containing new SM-charged states near the UV scale $\Lambda \sim M$. Importantly, such completions predict a host of additional phenomena that, while being model-dependent, are not captured by the low-energy EFT. Particularly for light scalars with sizable effective Yukawa couplings, $g_u \equiv c_Sv/(\sqrt{2}M)$, the new states cannot be too far above the weak scale, leading to additional experimental constraints and opportunities. The goal of this work is to investigate these issues within the context of two simple completions, one involving a VLQ and another with a second scalar doublet. For concreteness we focus on completions of the up quark-specific couplings. 

A second, more challenging question concerns the UV origin of the flavor-specific coupling structure. It should be stressed that the symmetry breaking pattern in Eq.~(\ref{eq:cS-break}) is a hypothesis on the form of the low energy EFT. As discussed Ref.~\cite{Batell:2017kty}, this assumption is self-consistent in that there are no large radiatively generated deviations from the flavor-specific structure, but its UV origin remains obscure. We do not endeavor here to construct explicit flavor models that naturally enforce flavor-specific couplings, but leave this important open question to future work. See also  Refs.~\cite{Knapen:2015hia,Altmannshofer:2017uvs} for some potential model-building approaches along this direction. 

Flavor-specific scalars may have any number of phenomenological applications, including as a possible new physics explanation for certain experimental anomalies (e.g., the muon anomalous magnetic moment discrepancy~\cite{Batell:2017kty}) or as a mediator between the SM and a dark sector. The latter application was considered in detail in Ref.~\cite{Batell:2018fqo}, which studied a light sub-GeV ``hadrophilic'' dark sector consisting of a Dirac fermion dark matter, $\chi$, coupled to an up quark-specific scalar mediator. Restricting ourselves here to real couplings for simplicity, the dominant low energy interactions in this scenario are
\begin{equation}
\label{eq:S-couplings}
{\cal L} \supset  -g_u S\bar u u - g_\chi S \overline \chi \chi,
\end{equation}
where the effective scalar-up quark coupling $g_u$ originates from the dimension-five operator in Eq.~(\ref{eq:bsmlag}),
\be
\label{eq:gu-coupling}
g_u\equiv\frac{c_S\, v}{\sqrt2 M},
\ee
with $v = 246$ GeV being the SM Higgs vacuum expectation value (vev). Through these couplings, the dark matter can obtain the correct relic abundance via thermal freeze-out of its annihilation either directly to hadrons or to scalar mediators. This scenario presents a rich low energy phenomenology, both for the case of visible scalar decays to hadrons (or photons if $m_S < 2 m_\pi$) and the case of invisible decays of scalars to dark matter particles. As we will demonstrate below in Secs.~\ref{sec:VLQ-completion} and \ref{sec:SD-completion}, the additional signatures predicted by the specific UV completions studied in this work can provide complementary constraints on this parameter space.

Starting from the EFT~(\ref{eq:bsmlag}) defined at the UV scale $M$, one can estimate the expected radiative size of other couplings in the EFT, which has implications for the naturalness of the light singlet scalar and its phenomenology. Concerning naturalness, for example, the two loop correction to the scalar mass and the shift to the up quark mass generated by the $S$ vev are small provided 
\begin{align}
\label{eq:natural-criterion}
c_S &\lesssim (16 \pi^2)\, \frac{m_S}{M} \approx 0.08 \left(\frac{m_S}{1 \, {\rm GeV}}\right) \left( \frac{2 \, {\rm TeV}}{M}\right), \nonumber \\ \Longrightarrow ~~ g_u & \lesssim \frac{16 \pi^2}{\sqrt{2}}\, \frac{m_S v}{M^2} \approx 0.007 \left(\frac{m_S}{1 \, {\rm GeV}}\right) \left( \frac{2 \, {\rm TeV}}{M}\right)^2.
\end{align}

As another example, there can be new loop-level contributions to FCNCs in the EFT. Considering the case of neutral kaon mixing, we find a three loop contribution described by the effective lagrangian
\begin{equation}
\label{eq:kaon}
{\cal L} \supset C^{ds} [ \overline d_L\gamma^\mu  s_L][ \overline d_L \gamma^\mu s_L]+{\rm h.c.},
\end{equation}
where the Wilson coefficient is estimated to be
\begin{equation}
C^{ds} \sim\frac{|c_S|^4 (V_{ud}^* V_{us})^2}{(16 \pi^2)^3M^2}.
\end{equation}
The current bound on this coupling is given by  ${\rm Re}[C^{ds}] \lesssim (10^3 \, {\rm TeV})^{-2}$~\cite{Bona:2007vi}, leading to a rather mild constraint 
\begin{equation}
\label{eq:EFT-kaon-bound}
    c_S \lesssim 4 \left(\frac{M}{2 \, {\rm TeV}}\right)^{1/2} ~~\Longrightarrow ~~ g_u \lesssim 0.4   \left(\frac{M}{2 \, {\rm TeV}}\right)^{-1/2}.
\end{equation}
As we will see, corrections to the scalar mass and kaon mixing operators arise already at one loop in the UV completions we study, which can lead to stronger conditions than shown in Eqs.~(\ref{eq:natural-criterion},\ref{eq:EFT-kaon-bound}). These examples highlight how the UV theory can provide complementary information on the theoretically favored or experimentally allowed model parameter space. 

With this introduction, in the next sections, we will analyze renormalizable models that lead to the low energy EFT in Eq.~(\ref{eq:bsmlag}), focusing on the case of the up-specific hypothesis for concreteness. Two simple completions of the dimension-five operator involve a new VLQ or scalar doublet at the scale $M$. For each of these possibilities, we will study the implications of the new high-scale physics for the radiatively generated corrections to the Lagrangian, as well as for phenomenology. We will find that naturalness and experimental constraints on the UV theories are in some cases stronger than in the the effective theory and probe complementary regions of low energy scalar mass--coupling parameter space. This suggests that only considering limits in the EFT does not provide a complete picture of the status of flavor-specific scalar theories.

%%%%%%%%%%%%%%%%%%%%%%%%%%%%%%%%%%%%%%%%
%%%%%%%%%%%%%%%%%%%%%%%%%%%%%%%%%%%%%%%%
\section{Vector-like Quark Completion}
\label{sec:VLQ-completion}

In this section we consider a renormalizable completion of the flavor-specific EFT in Eq.~(\ref{eq:bsmlag}) involving a VLQ. In what follows, we begin by presenting the model and then consider the natural expected radiative size of the scalar potential and other couplings in the theory, which will lead to a set of naturalness criteria. Following this, we discuss the transition to the physical basis including the interactions and decays of the VLQ. We then study the phenomenology of the model, including the impact of CKM unitarity, FCNCs, EWPTs, CP violation, and searches at the LHC. At the end of this section we present a summary of these constraints and also illustrate how these bounds probe the low-energy EFT parameter space of a light up-philic scalar. 

%%%%%%%%%%%%%%%%%%%%%%%%%%%%%%%%%%%%%%%%
\subsection{Model}
\label{sec:VLQ-model}

We add to the SM a real gauge singlet scalar $S$ and a VLQ with the same quantum numbers as the SM right-handed up quark, $U'_{L,R} \sim ({\bf 3}, {\bf 1}, \tfrac{2}{3})$. The Lagrangian of the model is 
\begin{align}
\label{eq:L-VLQ-UR}
\Lagr_{\rm VLQ} = \Lagr_{\rm SM} &+ \frac{1}{2} \partial_\mu S \partial^\mu S - \frac{1}{2} m_S^2 S^2
+\overline U' i \gamma^\mu D_\mu U' - M \, \overline U' U' \\
&
- [\, y_i \, \overline Q_L^{\, i} U'_R \, H_c 
+ \lambda^i \, \overline U'_L u_{R\, i} \,S   +{\rm h.c.}\,] \nonumber
 \end{align}
Here $i = 1,2,3$ is a generation index and $M$ is the VLQ mass. Integrating out the VLQ leads to an effective Lagrangian, with the leading terms appearing at the dimension 5 level:
\begin{equation}
\label{eq:L5}
{\cal L} \supset    \frac{y_i \, \lambda^{j}}{M} S \, \overline Q_L^{\, i} \, u_{R\,j} \, H_c +{\rm h.c.}
\end{equation}
Comparing this with the Wilson coefficient of the effective operator in Eq.~(\ref{eq:bsmlag}), we thus identify the VLQ mass $M$ as the new UV physics scale and $(c_S)_i^j \equiv - y_i \, \lambda^j$. The up-specific hypothesis corresponds to $y_i \propto \delta_{i1}$ and $\lambda^i \propto \delta^{i1}$ in the quark flavor basis in which $Y_u$ is diagonal.

It is important to note that the new physics couplings in Eq.~(\ref{eq:L-VLQ-UR}) are not the most general ones allowed by the gauge symmetries. To realize the flavor-specific hypothesis in the low-energy EFT, an extended flavor hypothesis must be made in the renormalizable completion. This entails specifying the spurion quantum numbers of $Y_u$, $Y_d$, $y$, $\lambda$, and $M$ under the enlarged quark flavor symmetry and how their background values break this symmetry. Once this hypothesis is made, the Lagrangian in the basis (\ref{eq:L-VLQ-UR}) is obtained through suitable quark flavor rotations. The flavor hypothesis for the VLQ completion is described in detail in Appendix~\ref{app:VLQ-flavor-hypothesis}.

In addition, the symmetries of the model admit additional renormalizable terms beyond those listed in Eq.~(\ref{eq:L-VLQ-UR}), such as a $S\overline UU$ Yukawa couplings, $S$ self-couplings, and interactions between $S$ and $H$. For simplicity, we assume that these are small, comparable to their radiatively induced contributions (see below) which provide a rough lower bound on the sizes of these couplings in the absence of fine-tuning.

%%%%%%%%%%%%%%%%%%%%%%%%%%%%%%%%%%%%%%%%
\subsection{Naturalness considerations}
\label{sec:VLQ-naturalness}

We are interested in the phenomenology of a light singlet scalar, $m_S \ll v$, with sizable couplings to the up quark. To achieve this, the UV model couplings $y$, $\lambda$ in Eq.~(\ref{eq:L-VLQ-UR}) must not be too small given the expectation that $M \sim {\cal O}({\rm TeV})$. However, it is of interest to know if the required magnitudes of these and other couplings in the theory are technically natural, i.e., that radiatively induced corrections to the Lagrangian parameters in \eqref{eq:L-VLQ-UR} are comparable to or smaller than the physical values of these parameters.

Since we are only interested in order-of-magnitude naturalness ``bounds'', we estimate the size of the loop corrections by using a factor $(16\pi^2)^{-1}$ for each loop and counting the relevant coupling and scale factors. For the latter, all mass scales that are parametrically smaller than $M$ can be neglected (such as all SM masses).

The most important corrections are those to the scalar masses, which arise at one loop in the renormalizable VLQ completion. In particular, the coupling $\lambda$ leads to a correction to the $S$ mass, $\delta m_S^2 \sim {\rm Tr}( \lambda^* \lambda)M^2/16 \pi^2$, where we have defined the matrix $(\lambda^* \lambda)^i_j =  \lambda^*_i \lambda^j $ and its trace ${\rm Tr} \, \lambda^* \lambda = \lambda^*_i \lambda^i $. Demanding this is less than the $S$ squared mass leads to the condition 
\begin{equation}
\label{eq:lambda-bound}
\lambda^i \lesssim 4 \pi \frac{m_S}{M} \simeq  (6 \times 10^{-3})   \left(  \frac{m_S}{1\, \rm GeV}  \right)  \left(  \frac{2 \, \rm TeV}{M}  \right).
\end{equation}
In addition, there is a correction to the Higgs mass term originating from the $y$ coupling, $\delta m_H^2 \sim {\rm Tr}(y y^*)M^2/16 \pi^2 $. Requiring that this is smaller than the square of the electroweak vev gives the naturalness condition
\begin{equation}
\label{eq:y-bound}
y_i \lesssim 4 \pi \frac{v}{M} \simeq  2  \left(  \frac{2 \, \rm TeV}{M}  \right).
\end{equation}
Combining Eqs.~(\ref{eq:lambda-bound}) and (\ref{eq:y-bound}) we obtain a bound on the Wilson coefficient $c_S$ defined in Eqs.~(\ref{eq:bsmlag},\ref{eq:L5}):
\begin{equation}
\label{eq:cS-bound-y-lambda}
 (c_S)_i^j 
 \lesssim 16 \pi^2 \,  \frac{v \, m_S}{M^2} \simeq   0.01  \left( \frac{m_S}{1\, \rm GeV}  \right)  \left(  \frac{2 \, \rm TeV}{M}  \right)^2.
\end{equation}
We note that this condition is stronger than the one obtained in the EFT, Eq.~(\ref{eq:natural-criterion}), by a factor $v/M$. Eq.~(\ref{eq:cS-bound-y-lambda}) confirms the general expectation that a light scalar with substantial couplings is in tension with naturalness considerations.

The Higgs portal operator $S^2 |H|^2$ will also give a correction to the $S$ mass term after electroweak symmetry breaking. The radiative size of this operator is estimated to be $\delta_{S^2 H^2}  \sim  {\rm Tr}[(y \lambda) (y \lambda)^\dag]/16 \pi^2 =  {\rm Tr}( c_S c_S^\dag) / 16 \pi^2$, and the correction to the scalar mass is thus $\delta m_S^2 \sim {\rm Tr}(c_S c_S^\dag) v^2 / 16 \pi^2 $.  The corresponding naturalness bound is thus
$(c_S)_i^j  \lesssim 4 \pi \,  m_S /v $, which is a weaker bound than Eq.~(\ref{eq:cS-bound-y-lambda}) so long as $M \gtrsim 2 \sqrt{\pi} v \sim $ TeV.

Besides the scalar masses, there are other corrections to the scalar potential that must be taken into account. In particular, there is an $S$  tadpole generated at two loops with size $\delta_S \sim 
{\rm Tr}( y \lambda \, Y_u^\dag ) M^3 / (16 \pi^2)^2  
=  {\rm Tr}( c_S \, Y_u^\dag)M^3/(16 \pi^2)^2$ . Provided the naturalness bounds in Eqs.~(\ref{eq:lambda-bound},\ref{eq:y-bound}) are satisfied, it is straightforward to show that the tadpole and mass terms dominate the $S$ potential; for a detailed argument in the EFT context, see Ref.~\cite{Batell:2017kty}. In the presence of the tadpole, the scalar develops a vev of characteristic size $v_S \simeq \delta_S/m_S^2 = {\rm Tr}( c_S \, Y_u^\dag) M^3/ (16 \pi^2)^2 m_S^2 , $ which in turn gives an effective contribution to the up quark Yukawa through the effective operator in Eq.~(\ref{eq:L5}) equal to $(\delta Y_u)_i^j  \simeq  (c_S)_i^j  \, {\rm Tr}(  c_S \, Y_u^\dag) M^2 /(16 \pi^2)^2 m_S^2 $ . Specializing to the flavor-specific hypothesis and demanding this correction is small compared to the SM Yukawa yields another naturalness condition, $c_S \lesssim 16 \pi^2 \, m_S/M $. This bound is clearly weaker than the one given in Eq.~(\ref{eq:cS-bound-y-lambda}). 

The other corrections to the scalar potential terms, such as the cubic interactions, $S^3$ and $S|H|^2$, and the quartic interactions $S^4$ and $|H|^4$, can be estimated in a similar manner. In particular, we note that $S|H|^2$ will induce mass mixing between the Higgs and the singlet scalars. However, as already mentioned, it can easily be seen that the expected radiative sizes of these couplings and the resulting Higgs-scalar mixing angle are tiny once the naturalness conditions~(\ref{eq:lambda-bound},\ref{eq:y-bound}) are met, and as such they will not play a role in our phenomenological considerations below. 

Besides the scalar potential, there are other couplings involving the quarks and scalar that are radiatively generated. The parametric dependence of the radiative sizes of these terms on the tree-level couplings follows from symmetry considerations~\cite{Batell:2017kty}. For instance, at one loop a mass mixing term  between the VLQ and SM up quark of the form ${\cal L}  \supset - m\, \overline U'_L  \,  u_R + {\rm h.c.}$ is generated with an expected radiative size $ m \sim y \, Y_u  \, M /16 \pi^2$. This is smaller than 1 MeV for  $y = 1$,  $Y_u \sim 10^{-5}$, and $M = 2 $ TeV. Therefore, no large tuning of the physical up quark mass is caused by this effect. Similarly, at one loop the coupling ${\cal L}  \supset -\lambda' \, \overline U'_L  U'_{R} \,S + {\rm h.c.}$ is generated with size $\lambda' \sim  y \lambda Y_u/16 \pi^2$, which is tiny if the naturalness bounds discussed above hold.

Given the considerations above, the dominant naturalness constraints come from the conditions on $y$ and $\lambda$ given in Eq.~(\ref{eq:lambda-bound},\ref{eq:y-bound}), which taken together  lead to the bound on $c_S$ given in Eq.~(\ref{eq:cS-bound-y-lambda}). 

%%%%%%%%%%%%%%%%%%%%%%%%%%%%%%%%%%%%%%%%
\subsection{Mixing, mass eigenstates, and interactions}
\label{sec:VLQ-mixing}

We now discuss the fermion mass diagonalization and the resulting interactions in the physical basis that will play an important role in our phenomenological considerations below. We start from the interactions of the VLQ, Eq.~(\ref{eq:L-VLQ-UR}), and the SM Yukawa couplings, Eq.~(\ref{eq:smlag}). Without loss of generality we may start from the flavor basis in which $Y_u$ is real and diagonal. Furthermore, invoking the up-specific hypothesis, the couplings $y$ and $\lambda$ in Eq.~(\ref{eq:L-VLQ-UR}) take the form $y_i = y \, \delta_{i1}$, $\lambda^i =  \lambda  \, \delta^{i1}$ in this basis. After electroweak symmetry breaking, there is mass mixing  between the $u$ and $U'$ quark fields,  
\begin{equation}
\label{massmixing-u-U}
-\Lagr=\left(\begin{array}{cc}
\overline {u}_{L} & \overline {U}'_{L}\end{array}\right)\left(\begin{array}{cc}
\displaystyle{\frac{y_u  v}{\sqrt{2}}} & \displaystyle{\frac{y v}{\sqrt{2}}}\\
\lambda v_S & M
\end{array}\right)\left(\begin{array}{c}
{u}_{R}\\
{U}'_{R}
\end{array}\right) +{\rm h.c.}
\end{equation}
where $y$, $\lambda$ and $M$ are complex parameters in general, while $y_u$ is real and positive in this basis. Through suitable phase rotations of the quark fields, it can be shown that there is one new physical phase if all of $y_u$, $y$, $\lambda$, and $M$ are non-vanishing. In the limit that any one of these couplings is zero, the phase can be rotated away. In Appendix~\ref{app:VLQ-scalar-couplings}, we provide a treatment of the diagonalization of the system (\ref{massmixing-u-U}) in the case when these four couplings take general values, as well as expressions for the quark interactions with electroweak gauge bosons and scalar bosons. Here, we instead consider the limit $y_u v, \lambda v_S \ll y v < M$, which is motivated by the fact that $y_u \ll y$ and the naturalness considerations regarding $y, \lambda, v_S$ discussed in Sec.~\ref{sec:VLQ-naturalness}. In this regime the system is diagonalized by a rotation of the left handed quarks,
\begin{align}
& {u}_{L}\rightarrow \cos\theta \,  {u}_{L} +\sin\theta \, {U}'_{L}, ~~~~~~~  {U}'_{L}\rightarrow \cos\theta \, {U}'_{L}-\sin\theta \, {u}_{L}, \label{eq:uUmix} \\
&~~~~~~ \cos\theta=\frac{M}{m_{U'}}, ~~~~~~~~~~~~~~~~  \sin\theta=\frac{y v}{\sqrt{2}\,m_{U'}} .  \nonumber
\end{align}
where $m_{U'} =\sqrt{M^{2}+ y^2 v^2 /2 }$ is the physical mass of the heavy VLQ. 

This mixing plays an important role in VLQ phenomenology due to the modifications of the SM interactions and the couplings induced between the VLQ and light SM fields. For example, the $W$ boson couplings involving the SM up quark and VLQ are 
\begin{equation}
{\cal L} \supset \frac{g}{\sqrt{2}} W_\mu^+ \left(  \cos\theta \, V_{1i}  \, \overline u_L \gamma^\mu d_{Li}  + \sin\theta \, V_{1i} \, \overline U'_L \gamma^\mu d_{Li}     \right)  + {\rm h.c.},
\label{eq:VLQ-W}
\end{equation}
where $V$ is unitary and $i = 1,2,3$ runs over the three SM generations. The first term implies that the effective SM CKM matrix is no longer unitary, while the second term leads to the decay $U' \rightarrow d_i W^+$. Furthermore, the $Z$ boson couplings involving the up quark and VLQ include
\begin{equation}
{\cal L} \supset g_{uL} Z_\mu \, \overline u_L \gamma^\mu u_L +  \left(  \frac{g}{2c_W} \sin\theta \cos\theta \, Z_\mu \, \overline u_L \gamma^\mu U'_L 
   +{\rm h.c.}  \right).
   \label{eq:VLQ-Z}
\end{equation} 
The $Z$ coupling to left-handed up quarks $g_{uL}$ is shifted from its SM value as a result of $u$--$U'$ mixing, while the right-handed up quark coupling $g_{uR}$ is unaffected by this mixing: 
\begin{align}
\label{eq:Zuu}
g_{uL} &= g_{uL}^{\rm SM} + \delta g_{uL},
\qquad
\delta g_{uL} = \sin^2\theta(g_{uR}^{\rm SM}-g_{uL}^{\rm SM})
\approx \frac{y^2v^2}{2M^2}(g_{uR}^{\rm SM}-g_{uL}^{\rm SM}) \\
g_{uL}^{\rm SM} &= \frac{g}{c_W}\Bigl(\frac{1}{2}-\frac{2}{3}s_W^2\Bigr),
\qquad
g_{uR}^{\rm SM} = -\frac{2gs_W^2}{3c_W}. \nonumber
\end{align}
As we will discuss below, such shifts can be probed by electroweak precision tests. Furthermore, the second term in Eq.~(\ref{eq:VLQ-Z}) above leads to the decay $U' \rightarrow u Z$. Finally, there are interactions between the scalars and quarks, the most important of which are
\begin{equation}
-{\cal L} \supset \cos\theta \frac{y}{\sqrt{2}} \, h \, \overline  u_L \, U'_R - \sin\theta  \, \lambda \, S \, \overline  u_L \, u_R +  \cos\theta \,  \lambda \, S \, \overline  U'_L \, u_R + {\rm h.c.}  
\label{eq:VLQ-S}
\end{equation}
The first and third terms above lead to the VLQ decays $U' \rightarrow u h$ and $U' \rightarrow u S$, respectively. The second term is the induced coupling of $S$ to up quarks, which in the limit of large $M$ reproduces the EFT result discussed earlier in Eqs.~(\ref{eq:S-couplings},\ref{eq:gu-coupling}). 

\subsubsection{VLQ and singlet scalar decays }

From the couplings of $U'$ to vector and scalar bosons given above, Eqs.~(\ref{eq:VLQ-W},\ref{eq:VLQ-Z},\ref{eq:VLQ-S}), we obtain the partial decay widths of the VLQ:
\begin{align}
\label{eq:VLQ-uS-decay}
\Gamma(U' \rightarrow  u S) & =\cos^2\!\theta \, \frac{ \lambda^2  \, m_{U'} }{32 \, \pi} \left( 1- \frac{m_S^2}{m_{U'}^2}  \right)^2 \simeq \frac{ \lambda^2   M }{32 \, \pi} , \\
\label{eq:VLQ-uh-decay}
\Gamma(U' \rightarrow  u h) & = \sin^2\!\theta  \cos^2\!\theta \,\frac{ G_F \, m_{U'}^3}{16\sqrt{2} \, \pi} \left( 1- \frac{m_h^2}{m_{U'}^2}  \right)^2 \simeq \frac{ y^2  M }{64 \, \pi} , \\
\label{eq:VLQ-uZ-decay}
\Gamma(U' \rightarrow  u Z) & = \sin^2\!\theta \cos^2\!\theta \, \frac{   G_F \, m_{U'}^3}{16\sqrt{2}\, \pi} \left( 1- \frac{m_Z^2}{m_{U'}^2}  \right)^2
\left( 1+ \frac{2 m_Z^2}{m_{U'}^2}  \right) \simeq \frac{ y^2  M }{64 \, \pi} , \\
\label{eq:VLQ-dW-decay}
\Gamma(U' \rightarrow  d W) & = \sin^2\!\theta  \, \frac{   G_F \, m_{U'}^3}{8\sqrt{2} \,\pi} \left( 1- \frac{m_W^2}{m_{U'}^2}  \right)^2 \left( 1+ \frac{2 m_W^2}{m_{U'}^2}  \right) \simeq \frac{ y^2  M }{32 \, \pi},
\end{align}
where the $u$--$U'$ mixing angle $\theta$ is defined in Eq.~(\ref{eq:uUmix}). We have also provided approximate expressions for the decay widths in the limit $M \gg v$, from which is it is evident that the $U'$ decays to electroweak bosons respect the Goldstone Equivalence Theorem. Given the naturalness considerations discussed earlier, which suggest $\lambda \ll y$, we typically expect the $U'$ decays to electroweak bosons to dominate. As we will discuss in detail below, this suggests that LHC searches for VLQs with couplings to the first generation are a promising way to test this completion. 

However, it is also possible in principle that $U'$ could dominantly decay to a scalar $S$ and an up quark, provided $y \lesssim \lambda$. In such a situation, the VLQ signature will depend in detail on how $S$ decays. If there are no additional light states present in the theory, $S$ will decay to pairs of up quarks [or to exclusive hadronic modes for $m_S \sim {\cal O}( 1$ GeV)]. This decay width is controlled by the effective scalar-up quark coupling $g_u$ defined in Eqs.~(\ref{eq:S-couplings},\ref{eq:gu-coupling}). If $S$ is even lighter, with mass below the two-pion threshold, it will decay to a pair of photons at one loop, and is naturally long-lived. Alternatively, if there are additional light degrees of freedom with sizable couplings to $S$, the scalar may dominantly decay to such states. For example, in the case of a coupling to light dark matter as in Eq.~(\ref{eq:S-couplings}), the scalar can decay invisibly via $S \rightarrow \chi \overline \chi$. 

We now turn to the phenomenology of the model.

%%%%%%%%%%%%%%%%%%%%%%%%%%%%%%%%%%%%%%%%
\subsection{CKM constraints}
\label{sec:VLQ-CKM}
Due to the mixing of the up quark with the VLQ, the effective $3\times3$ CKM matrix describing the mixing of the SM quarks is no longer unitary. This is clearly seen in Eq.~(\ref{eq:VLQ-W}), where the elements of the unitary matrix $V_{1i}$ are multiplied by the prefactor $\cos\theta$. This model therefore predicts that the top-row CKM unitarity triangle relation is modified and no longer equal to unity. The current experimental determination of the top-row CKM unitarity relation is~\cite{Zyla:2020zbs}
\begin{equation}
 \Big[ \, |\tilde{V}_{ud}|^{2}+|\tilde{V}_{us}|^{2}+|\tilde{V}_{ub}|^{2} \,\Big] 
\Big\vert_{\rm exp}= 0.9985(3)_{V_{ud}} (4)_{V_{us}},
\label{eq:top-row}
\end{equation}
where the dominant uncertainties from $V_{ud}$ and $V_{us}$ are indicated. Here $\tilde{V}_{ij}$ are the \emph{apparent} CKM matrix elements when assuming the SM. Interestingly, the current determination (\ref{eq:top-row}) displays a 3$\sigma$ deviation from unitarity. Such a deviation is a natural consequence of our model, which gives the prediction
\begin{align}
|\tilde{V}_{ud}|^{2}+|\tilde{V}_{us}|^{2}+|\tilde{V}_{ub}|^{2}
&= \cos^2\!\theta \, \Big[ \, |V_{ud}|^{2}+|V_{us}|^{2}+|V_{ub}|^{2} \, \Big] = \cos^2 \theta,
\end{align}
where we have used the unitarity of $V$ in the second step. This is to be compared with Eq.~(\ref{eq:top-row}). The model can therefore provide an explanation of this discrepancy provided the mixing angle is in the range
\begin{equation}
0.032 < |\sin \theta \,| < 0.045, 
\end{equation}
which brings the theory prediction and experimental determination into agreement at the 1$\sigma$ level. Explaining this discrepancy with VLQ was also recently studied in Refs.~\cite{Belfatto:2019swo,Belfatto:2021jhf,Branco:2021vhs}.

Beyond a possible explanation of this  discrepancy, Eq.~(\ref{eq:top-row}) can be used to place a conservative bound on the mixing angle. Requiring that the theory prediction is within 3$\sigma$ of the experimental determination, we find the constraint $\sin\theta \lesssim 0.055$, which can be phrased as the following bound on the model parameters using Eq.~(\ref{eq:uUmix}):
\begin{align}
y\lesssim 0.6\left(\frac{M}{2\,{ \rm TeV} }\right).
\label{eq:vlqckm}
\end{align}

%%%%%%%%%%%%%%%%%%%%%%%%%%%%%%%%%%%%%%%%
\subsection{FCNCs}

Although the flavor-specific hypothesis generally provides strong protection against FCNCs, there can still be important effects if the VLQ is light enough and its couplings are relatively large. Here we consider the contributions to neutral kaon mixing, which generally provides the strongest FCNC constraints. In particular, there is a one loop box diagram resulting from $U'$ and Higgs exchange, which leads to an effective operator with four $Q_L$ fields.  The resulting effective Lagrangian reads
\begin{equation}
{\cal L} \supset - \frac{y_i y^{\dag j} y_k y^{\dag \ell}}{128 \pi^2 M^2} [ \overline Q^{\, i} \gamma^\mu P_L \, Q_j][ \overline Q^{\, k} \gamma^\mu P_L \, Q_\ell].
\end{equation}
Going to the physical basis and specializing to the up-specific hypothesis, we find a contribution that mediates neutral kaon mixing, described by the effective Lagrangian (\ref{eq:kaon}) with the Wilson coefficient 
\begin{equation}
 C^{ds} =  
 - \frac{y^4 |V_{ud}^* V_{us}|^2}{128 \pi^2 M^2}. 
\end{equation}
Current limits restrict  ${\rm Re}[C^{ds}] \lesssim (10^3 \, {\rm TeV})^{-2}$~\cite{Bona:2007vi}, leading to the constraint
\begin{equation}
y \lesssim 0.6 \left(\frac{M}{2 \, {\rm TeV}}\right)^{1/2}.
\label{eq:vlqfcnc}
\end{equation}

%%%%%%%%%%%%%%%%%%%%%%%%%%%%%%%%%%%%%%%%
\subsection{Electroweak precision bounds}

%------------------------------
\begin{figure}
    \includegraphics[width=0.6\textwidth]{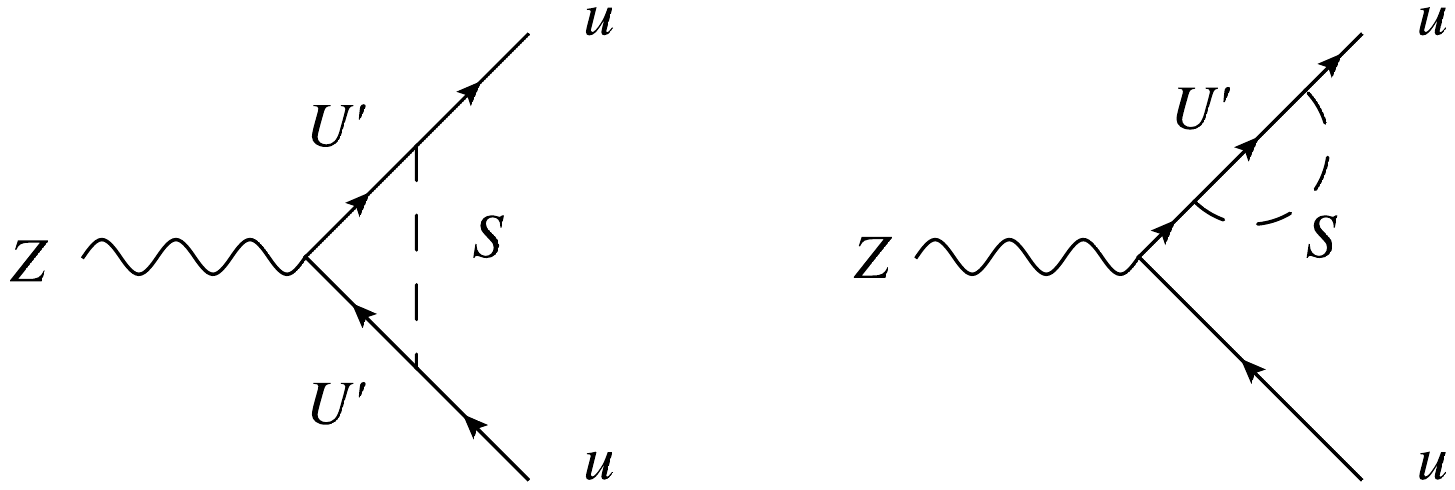} \\
   \makebox[0.3\textwidth]{\centering (a)} 
    \makebox[0.3\textwidth]{\centering (b)}
    \caption{Loop diagrams contributing to the hadronic $Z$ width in the VLQ  completion.
    }
    \label{fig:ewpQ}
\end{figure}
%------------------------------

The heavy VLQ modifies the partial width of $Z$ to hadrons in two ways: through $u$--$U'$ mixing and through the loop diagrams in Fig.~\ref{fig:ewpQ}~(a,b). Additional diagrams suppressed by both a loop factor and the mixing angle $\theta$ exist but will be neglected. The main observable to constrain modifications of the hadronic $Z$ width is the hadron-to-lepton branching ratio, $R_\ell \equiv \frac{\Gamma[Z \to \rm had.]}{\Gamma[Z \to \ell^+\ell^-]}$. The current experimental data and SM theory prediction give $R_\ell^{\rm exp} - R_\ell^{\rm SM} = 0.034\pm 0.025$ \cite{Tanabashi:2018oca}. For general shifts in the $Z$ boson coupling to up and down quarks, $\delta g_{uL,R}$, $\delta g_{dL,R}$, the modification to this observable is given by 
\begin{equation}
    \delta R_\ell \simeq \frac{2 N_c\, {\rm Re}( g^{\rm SM}_{uL} \, \delta g_{uL} + g^{\rm SM}_{uR} \,\delta g_{uR}+ g^{\rm SM}_{dL} \, \delta g_{dL} + g^{\rm SM}_{dR} \,\delta g_{dR}  )}{(g^{\rm SM}_{\ell L})^2 + (g^{\rm SM}_{\ell R})^2},
    \label{eq:dRl}
\end{equation}
where $N_c = 3$ and $g^{\rm SM}_{fL} = \tfrac{g}{c_W}(T_f^3 - Q_f s_W^2)$, $g^{\rm SM}_{fR} = \tfrac{g}{c_W}(- Q_f s_W^2)$ are the $Z$ boson couplings to fermions $f$ in the SM. 

The largest effect comes from the mixing in \eqref{eq:uUmix}, which leads to a tree-level shift of the $Z\bar{u}_L u_L$ coupling, given above in Eqs.~(\ref{eq:VLQ-Z},\ref{eq:Zuu}). Plugging these shifts into Eq.~(\ref{eq:dRl}) we obtain 
\begin{equation}
    \delta R_\ell \simeq -\frac{3(\tfrac{1}{2}-\tfrac{2}{3} s_W^2)}{(-\tfrac{1}{2}+ s_W^2)^2+(s_W^2)^2}\sin^2\theta \simeq -8.3 \sin^2\theta  
    \label{eq:dRl-tree}
\end{equation}
This leads to the bound 
\begin{equation}
|\sin\theta\,| \lesssim 0.044.
\label{eq:vlqewpt}
\end{equation}
For $v\ll M$, the bound can also be stated as  $|yv/M| < 0.063$.

This bound could be improved at a future high-luminosity $e^+e^-$ collider running on the $Z$-pole, such as CEPC \cite{CEPCStudyGroup:2018ghi} or FCC-ee \cite{Abada:2019zxq}. With the expected FCC-ee precision, $\delta R_\ell^{\rm exp.} = 0.001$ \cite{Blondel:2018mad}, one would be able to constrain $|yv/M| <  0.022$.

At the one-loop level, the diagrams Fig.~\ref{fig:ewpQ}~(a,b) generate a correction to the $Z\bar{u}_Ru_R$ coupling. In the limit $M \gg v \gg m_S$, it is given by
\be
\label{eq:ewpQ1l}
\delta g_{uR} \approx g_{uR}^{\rm SM}\, \frac{7\lambda^2}{576\pi^2}\,\frac{m_Z^2}{M^2}
\ee
For $M=1$~TeV and $\lambda \sim \sqrt{4\pi}$ near its perturbative limit, the shift in $R_\ell$ from \eqref{eq:ewpQ1l} is less than $10^{-3}$ and thus phenomenologically irrelevant.

%%%%%%%%%%%%%%%%%%%%%%%%%%%%%%%%%%%%%%%%
\subsection{CP violation}
\label{sec:VLQ-CPviolation}

If the couplings $M$, $y$, $\lambda$ are complex we may expect new CP-violating phenomena including a potentially large neutron electric dipole moment. Separate rephasings of $u_{L,R}$ and $U_{L,R}^\prime$ leave invariant
\begin{equation}
    \phi_{\rm CP}\equiv{\rm arg}\left[y_u M \left(y\lambda\right)^\ast\right],
\end{equation}
and all CP-violating effects are proportional to $\sin\phi_{\rm CP}$.
The dominant contribution in the VLQ model arises due to an effective CP-violating four up quark operator mediated by the exchange of the scalar $S$,
\begin{equation}
\label{eq:CPodd4up}
{\cal L }  \supset C'_u \, \overline u i \gamma^5  u \, \overline u u, 
\end{equation}
where the Wilson coefficient is
\begin{equation}
\label{eq:CPodd4upWilson}
C'_u = \frac{{\rm Re}(Y_{S \bar u u}) {\rm Im}(Y_{S \bar u u})   }{m_S^2} \simeq - \frac{y^2 \lambda^2 v^2}{4M^2 m_S^2 }\sin{2\phi_{\rm CP}}.
\end{equation}
The scalar-quark couplings are defined in the appendix, Eq.~(\ref{eq:S-quark-couplings}). The final expression in Eq.~(\ref{eq:CPodd4upWilson}) holds provided $y_u M \gg y \lambda v_S$, which is always satisfied in the natural region of parameter space. 
The effective operator, Eq.~(\ref{eq:CPodd4up}), is then matched to CP-violating interactions in the chiral Lagrangian, from which the relevant hadronic matrix elements can be estimated. For this we use the results of Ref.~\cite{An:2009zh}, which derives a prediction for the neutron EDM in terms of the Wilson coefficient,
\begin{equation}
\label{eq:neutronEDM}
d_n = 0.182 \, e \, C'_u \, {\rm GeV} \simeq 3.6 \times 10^{-15} \, e \, {\rm cm} \, C'_u \, {\rm GeV}^2.
\end{equation}
The current leading upper limit on the neutron EDM  is $|d_n| < 1.8 \times 10^{-26} e$ cm  (90$\%$ C.L.) from Ref.~\cite{Abel:2020gbr}. Using Eqs.~(\ref{eq:CPodd4upWilson},\ref{eq:neutronEDM}) we can express this as a limit on the effective coupling of the scalar to up quarks ($g_u \simeq y \lambda v /\sqrt{2} M$), 
\begin{equation}
\label{eq:neutronEDM4up}
|g_u|  \sqrt{\sin 2 \phi_{\rm CP}} < 3 \times 10^{-6}\left( \frac{m_S}{1 \, \rm GeV}\right)
\end{equation}

A one-loop contribution to the neutron EDM also arises due to pion-scalar mixing which leads to a CP-violating pion-nucleon coupling. The bound that results from this process is~\cite{Seng:2014pba}
\begin{equation}
\label{eq:neutronEDMspiloop}
|g_u|  \sqrt{\sin 2 \phi_{\rm CP}} < 1 \times 10^{-5}\left( \frac{m_S}{1 \, \rm GeV}\right),
\end{equation}
which is quantitatively similar to that in Eq.~(\ref{eq:neutronEDM4up}).\footnote{Note that Ref.~\cite{Batell:2017kty} also included an estimate of the one-loop contribution to the neutron EDM in the presence of pion-scalar mixing (see Eq. (35) in that reference) by matching to the chiral Lagrangian and cutting the loop off at the neutron mass. The resulting contribution to $d_n$ from this process in~\cite{Batell:2017kty} is larger by a factor $\left(2\frac{m_\pi^2}{m_N^2}\log\frac{m_N^2}{m_\pi^2}\right)^{-1}\simeq 6$ than that in Ref.~\cite{Seng:2014pba} which involves a detailed treatment of heavy baryon chiral perturbation theory.}

Other contributions to the neutron EDM are subdominant to the four up-quark CP odd operator ~(\ref{eq:CPodd4up}). 
For example, a one-loop penguin-type diagram with the scalar $S$ entering in the loop, gives a contribution to the up quark EDM of
\begin{align}
d_u & \simeq \frac{3 e Q_u}{32 \pi^2} \left|g_u\right|^2 \sin 2 \phi_{\rm CP} \frac{m_u  }{m_S^2}\left[1+\frac{4}{3} \log\left(\frac{\Lambda_{\rm IR}}{m_S}\right)\right],
\end{align}
where we have taken the large $M$ limit and $\Lambda_{\rm IR}\simeq 300~{\rm MeV}$ is an IR cutoff on the loop.
The neutron EDM induced by the up-quark EDM is  $d_n = 0.784(28) d_u$~\cite{Gupta:2018lvp}. We thus obtain a bound, 
\begin{equation}
|g_u|  \sqrt{\sin 2 \phi_{\rm CP}}  < 3.2 \times 10^{-4}\left( \frac{m_S}{1 \, \rm GeV}\right)
\label{eq:penguinlimit}
\end{equation}
which is significantly weaker than the one given in Eq.~(\ref{eq:neutronEDM4up}). A similar diagram leads to an up quark chromo-EDM, leading to a comparable limit to that in Eq.~(\ref{eq:penguinlimit}) from the mercury EDM limit of $\left|d_{\rm Hg}\right|<7.4\times 10^{-30}\,e\,{\rm cm}$~\cite{Graner:2016ses}.

In Fig.~\ref{fig:VLQlowenplots}, we show the leading limit on $g_u$ from Eq.~(\ref{eq:neutronEDM4up}) fixing $\phi_{\rm CP}=\pi/4$. Since the estimates in this section all assume that $m_S$ is larger than the hadronic scale, we only display this limit for $m_S>1~{\rm GeV}$.

%%%%%%%%%%%%%%%%%%%%%%%%%%%%%%%%%%%%%%%%
\subsection{Collider phenomenology}
\label{sec:VLQ-collider}

We now discuss the collider phenomenology of the VLQ completion. Pair production of $U'$ at hadron colliders proceeds through the strong interaction, while single electroweak production is also possible through mixing. Unlike top partners, the $U'$ decays only to light flavor quarks, so typical VLQ searches requiring $b$-tagged jets in the final state do not apply. Instead, we consider collider searches for VLQs decaying to light quarks. Motivated by the naturalness constraints on $\lambda$, Eq.~(\ref{eq:lambda-bound}), we will initially focus on the small $\lambda$ limit, where the $U' \to S u$ decay can be neglected and VLQ decays to a first generation quark and an electroweak boson dominates, see Eqs.~(\ref{eq:VLQ-uS-decay}-\ref{eq:VLQ-dW-decay}). The ATLAS and CMS collaborations performed light-flavor VLQ searches only with 8 TeV data to date. ATLAS considered pair production of $U'$ followed by the decay $U' \to W d$ in the single-lepton final state~\cite{Aad:2015tba}. CMS considered both pair production and single production, including the decay modes $U' \to W d, Z u, h u$ in final states involving one or more leptons~\cite{Sirunyan:2017lzl}. We will follow CMS, performing an analysis similar to their search for pair production of VLQs decaying to two leptons, jets and missing energy.

Before turning to our recast analysis, we briefly mention the other channels studied by CMS in Ref.~\cite{Sirunyan:2017lzl}. First, in principle both pair production and single production of the $U'$ is possible. However, single production requires mixing between the $U'$ and the SM quarks, which is strongly constrained. CMS searched for single production of down-type VLQ decaying to $W^- u$ or $Z d$. The latter decay mode is relevant to the present case of up-type VLQ, and in this channel the effective limit on the mixing angle is $\mathcal{O}(1)$ across the mass range considered. Since the single production cross-section goes as the square of the mixing angle and constraints from CKM and EW precision observables limit $\sin^2 \theta \lesssim 10^{-3}$, single production is not competitive with pair production in the allowed regions of parameter space. Turning to pair production, CMS performed searches for $U'\bar{U}'$ in single lepton, dilepton, and multilepton (3 or 4) final states. In the Goldstone equivalence limit where the ratio of the $U'$ decays to $W$, $Z$, and $h$ is 2:1:1 (see Eqs.~(\ref{eq:VLQ-uh-decay},\ref{eq:VLQ-uZ-decay},\ref{eq:VLQ-dW-decay})), the single lepton analysis is the strongest of these searches owing to the high $W$ branching fraction. However, this channel involves a kinematic fit of each event to the hypothesis that it contains two $W$ bosons, one $W$ and one $Z$, or one $W$ and one $h$. Events are considered under each hypothesis based on the $\chi^2$ of this fit, which is difficult to estimate. We thus choose to focus on the next most constraining channel, the dilepton final state. The multilepton search has much lower statistics.

For the signal, we simulate pair production of the $U'$ with MadGraph~\cite{Alwall:2014hca}, Pythia~\cite{Sjostrand:2014zea} and Delphes~\cite{deFavereau:2013fsa}, using the UFO~\cite{Alloul:2013bka,Degrande:2011ua} model for a singlet VLQ~\cite{Buchkremer:2013bha}. We also simulate the dominant backgrounds in the CMS search, which are top pair production and $Z$ + jets. We stay close to the cuts of the signal region aimed at the $WqWq$ final state, which enjoys the highest statistics due to the large $U' \to W d$ branching fraction. Specifically, we require:
\begin{itemize}
\item Exactly two opposite-sign leptons with $p_T > 30, 20~\mathrm{GeV}$ respectively and $|\eta| < 2.5$
\item At least two jets with $p_T > 200, 100~\mathrm{GeV}$ and $|\eta| < 2.4$ that do not pass a $b$-tag with efficiency 84\% and fake rate 10\%
\item No same-flavor lepton pair within 7.5 GeV of the $Z$ mass
\item Missing transverse energy (MET) > 60~GeV
\item $S_T$ > 1000~GeV, where $S_T$ is the scalar sum of the lepton $p_T$, jet $p_T$ and MET
\end{itemize}
Most of these cuts are very similar to those of CMS, except that while they set limits using the full $S_T$ distribution, we simply perform a cut-and-count analysis with a minimum $S_T$ requirement. Prior to this cut, our signal and background event counts are in agreement with CMS. We then estimate $2\sigma$ limits on the production cross-section as a function of $m_{U'}$, considering statistical uncertainties only.

\begin{figure}
\begin{center}
\includegraphics[width=1.0\textwidth]{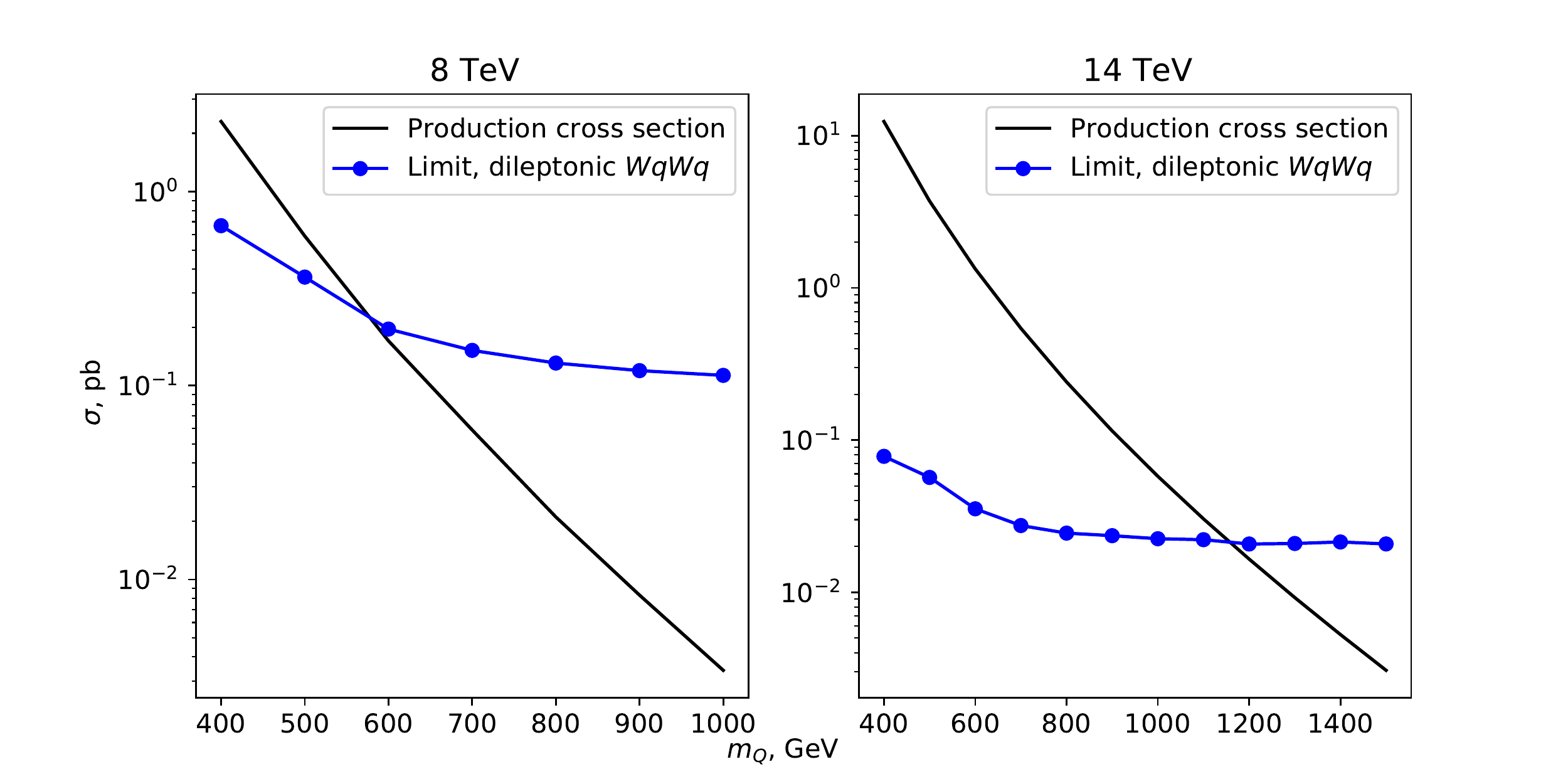} 
\end{center}
\caption{
Estimated limits on the $U'$ pair production cross-section from a search for a final state with two leptons, jets and missing energy. The analysis is close to that of one performed by CMS~\cite{Sirunyan:2017lzl}, and is shown assuming 20~fb$^{-1}$ of luminosity at 8~TeV (left) and 3000~fb$^{-1}$ of luminosity at 14~TeV (right).
}
\label{fig:vlqlight}
\end{figure}

We perform this search with 20~fb$^{-1}$ of integrated luminosity at 8~TeV as a check of our analysis, and then repeat it assuming 3000~fb$^{-1}$ at 14~TeV. Our results are shown in Figure~\ref{fig:vlqlight}. The expected 8 TeV limit on $m_{U'}$ is approximately 575~GeV. For comparison, CMS combines several dilepton and multilepton search channels to obtain a limit of 585~GeV when the branching fractions of $U'$ to $W d$, $Z u$, and $H u$ are 50\%, 20\%, and 30\%, respectively. At 14 TeV with the full HL-LHC dataset, we estimate that the limit from the dilepton channel alone could approach 1150~GeV. This represents a significant increase over the limit of 685~GeV reported by CMS in Ref.~\cite{Sirunyan:2017lzl} for a $U'$ which decays with the branching ratios expected by Goldstone equivalence, when combining searches in multiple pair production final states. It would be of interest, then, to see updated light-flavor VLQ searches with the latest LHC dataset. While we have considered only the dilepton final state, it is quite possible that a combination of searches, including the high statistics single-lepton channel, could do even better than our projection.

Next, we consider the case where the $U' \to S u$ decay is important. The relevant coupling $\lambda$ is limited by Eq.~(\ref{eq:lambda-bound}) if it is natural, which for light scalars $S$ is typically much smaller than the effective $\bar{Q} U' H_c$ coupling allowed by the indirect constraints from CKM unitarity, FCNCs and EWPT in Eqs.~(\ref{eq:vlqckm}), (\ref{eq:vlqfcnc}) and (\ref{eq:vlqewpt}) respectively. However, if $y$ is even smaller than required by these indirect constraints, the $U' \to S u$ decay could dominate. For visibly decaying $S$, the pions produced in the $S$ decay would be highly collimated if $S$ were light. Consequently, strong production of $U'$ could be probed by searches for pair production of dijet resonances. A reinterpretation~\cite{Easa:2020mfv} of a 13~TeV ATLAS paired dijet resonance search~\cite{Aaboud:2017nmi} found that for light $S$, the limit on the VLQ mass is approximately 700~GeV. For invisibly decaying $S$, searches for jets plus missing energy would apply, which tend to give considerably stronger limits~\cite{CMS:2019ybf,ATLAS:2020syg}.

Finally, the light scalar can also be directly produced in hadron collisions, but the bounds on the effective scalar-up quark coupling $g_u$ are generally quite weak. For visible $S$ decays there are constraints from di-jet+photon searches in the mass range 10 GeV $\lesssim m_s \lesssim 100$~GeV, which lead to a bound $g_u \lesssim 0.3$~\cite{CMS:2019xai}.
For invisible decays of $S$, one can look for a mono-jet signature. A bound $g_u \lesssim 0.1$ was derived previously in Ref.~\cite{Batell:2018fqo}.

%%%%%%%%%%%%%%%%%%%%%%%%%%%%%%%%%%%%%%%%
\subsection{Summary}
\label{sec:VLQ-summary}

%%%%%%%%%%%%%%%%%%%%
\begin{figure}
\centering
\includegraphics[width=0.7\textwidth]{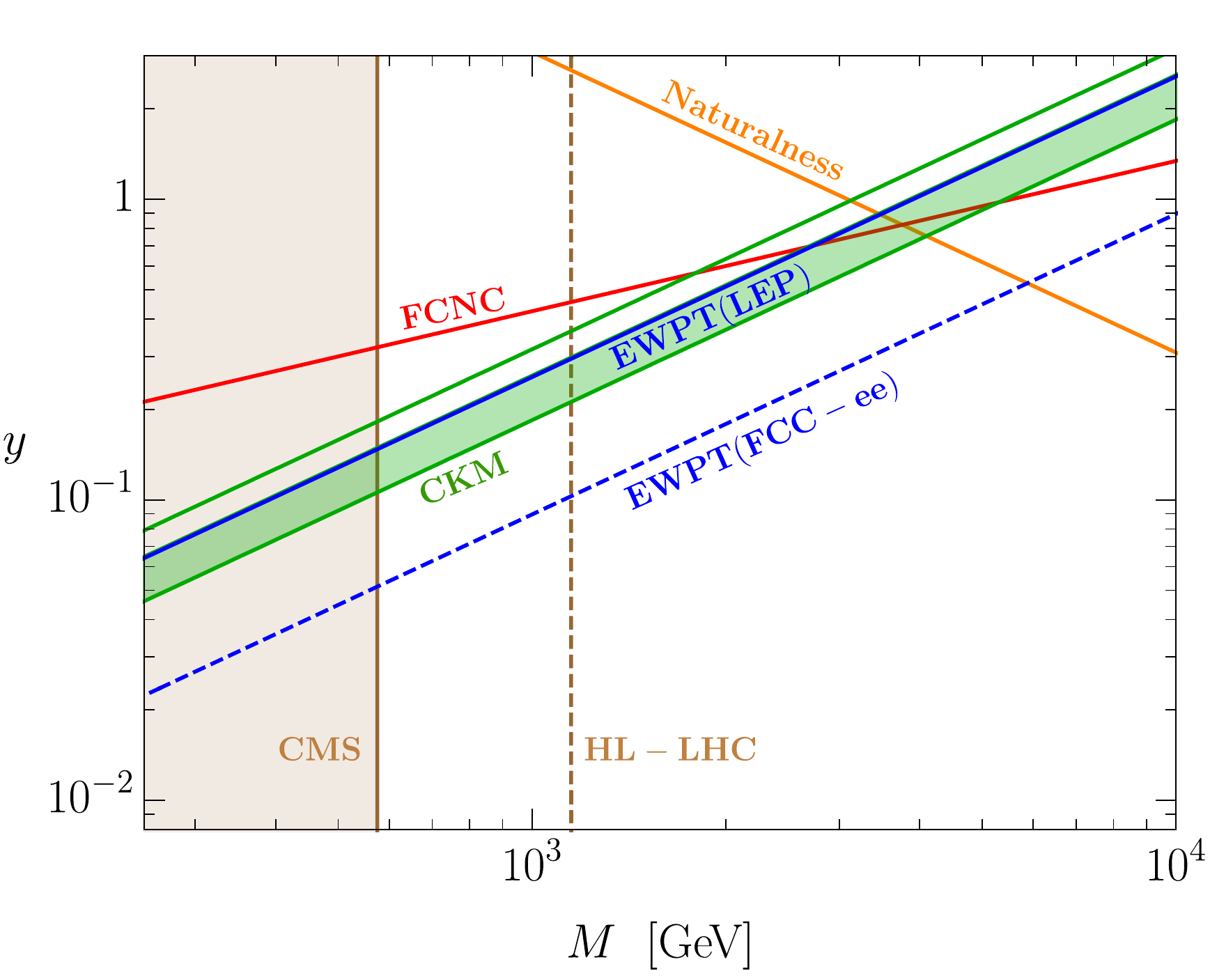}
\caption{\label{fig:VLQhighenplots}
Constraints on the VLQ model in the $M-y$ plane. Shown are current bounds from neutral kaon mixing (red solid line), CKM unitarity (green solid line), the $Z$ boson hadronic-to-leptonic branching ratio $R_\ell$ (blue solid line), and a direct VLQ search from CMS (brown shaded region). Regions above the lines are excluded. We also indicate the parameter space where the model can explain the $\sim 3 \sigma$ discrepancy in CKM top row unitarity triangle determination (green shaded band).
The expected future reach from precision measurements of $R_\ell$ at FCC-ee (blue dashed line) and a direct VLQ search at the HL-LHC (brown dashed line) are also indicated. Large couplings and VLQ masses do not satisfy the naturalness condition (\ref{eq:y-bound}) (orange solid line). This plot assumes $\lambda \ll y$, which is typically the case in this plane for light scalars, $m_S \lesssim$ GeV, and natural values of $\lambda$, as suggested by Eq.~(\ref{eq:lambda-bound}).
}
\end{figure}
%%%%%%%%%%%%%%

%%%%%%%%%%%%%%%%%%%%
\begin{figure}
\centering
\includegraphics[width=0.49\textwidth]{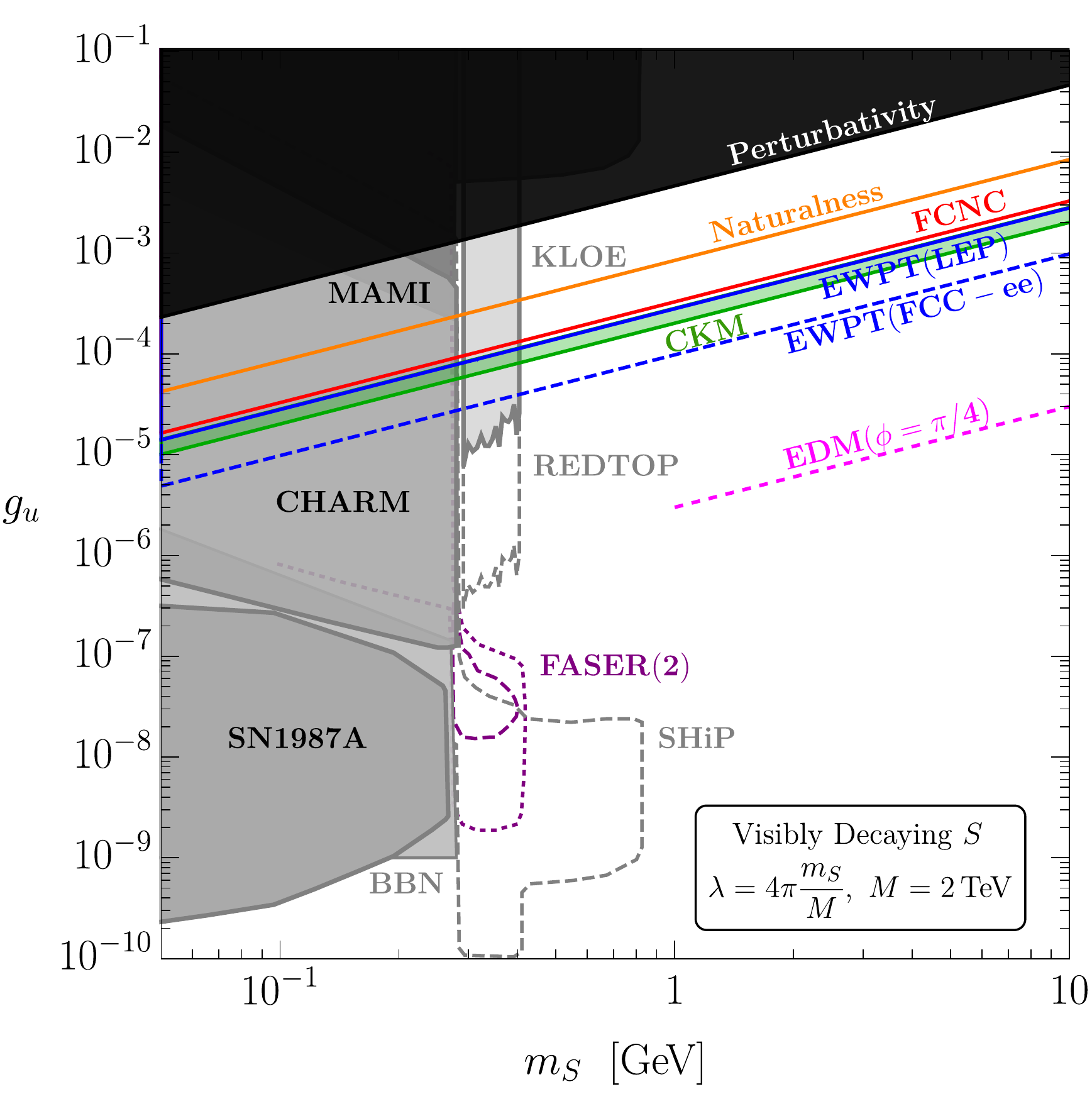}
\includegraphics[width=0.49\textwidth]{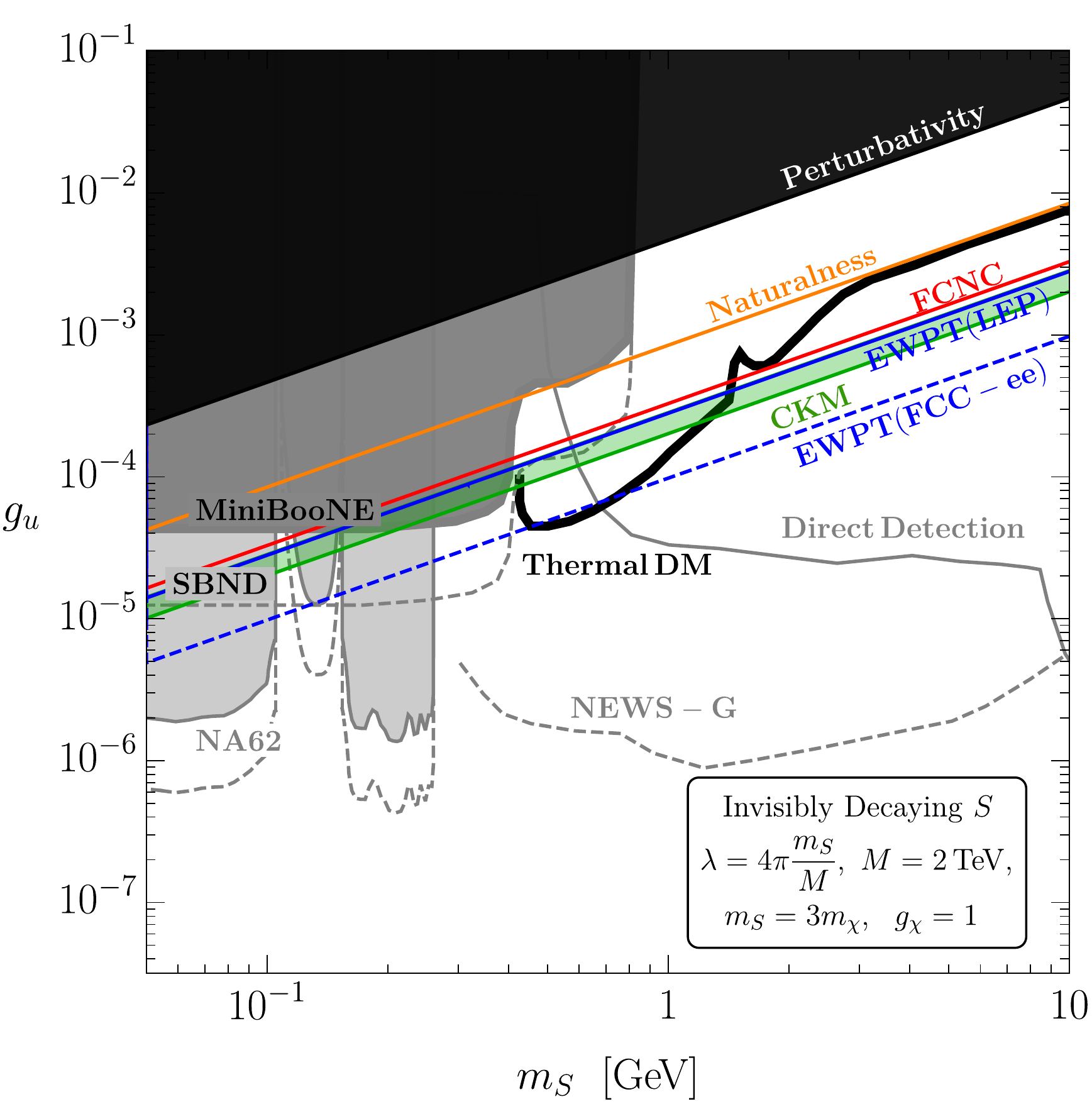}
\includegraphics[width=0.49\textwidth]{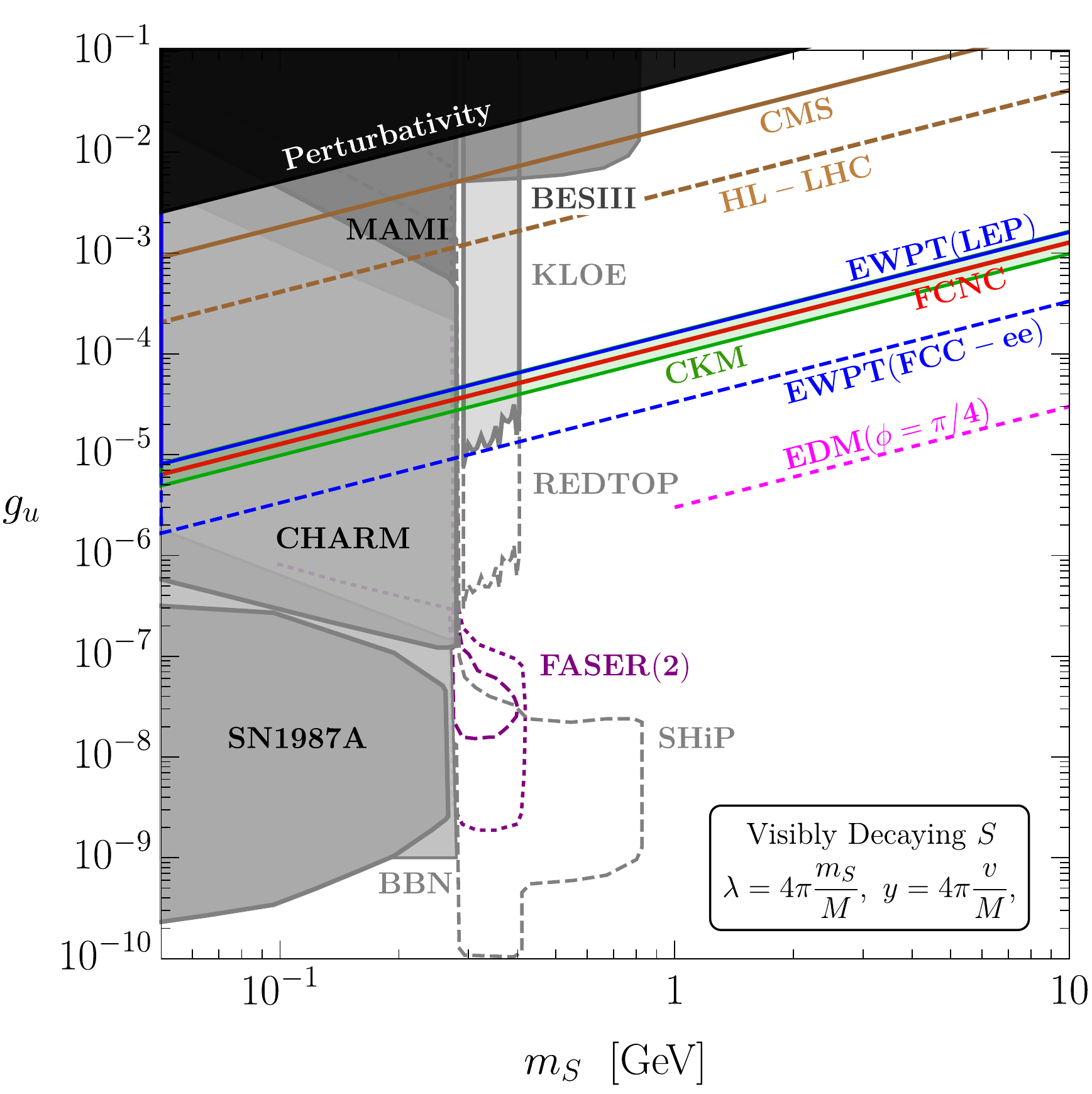}
\includegraphics[width=0.49\textwidth]{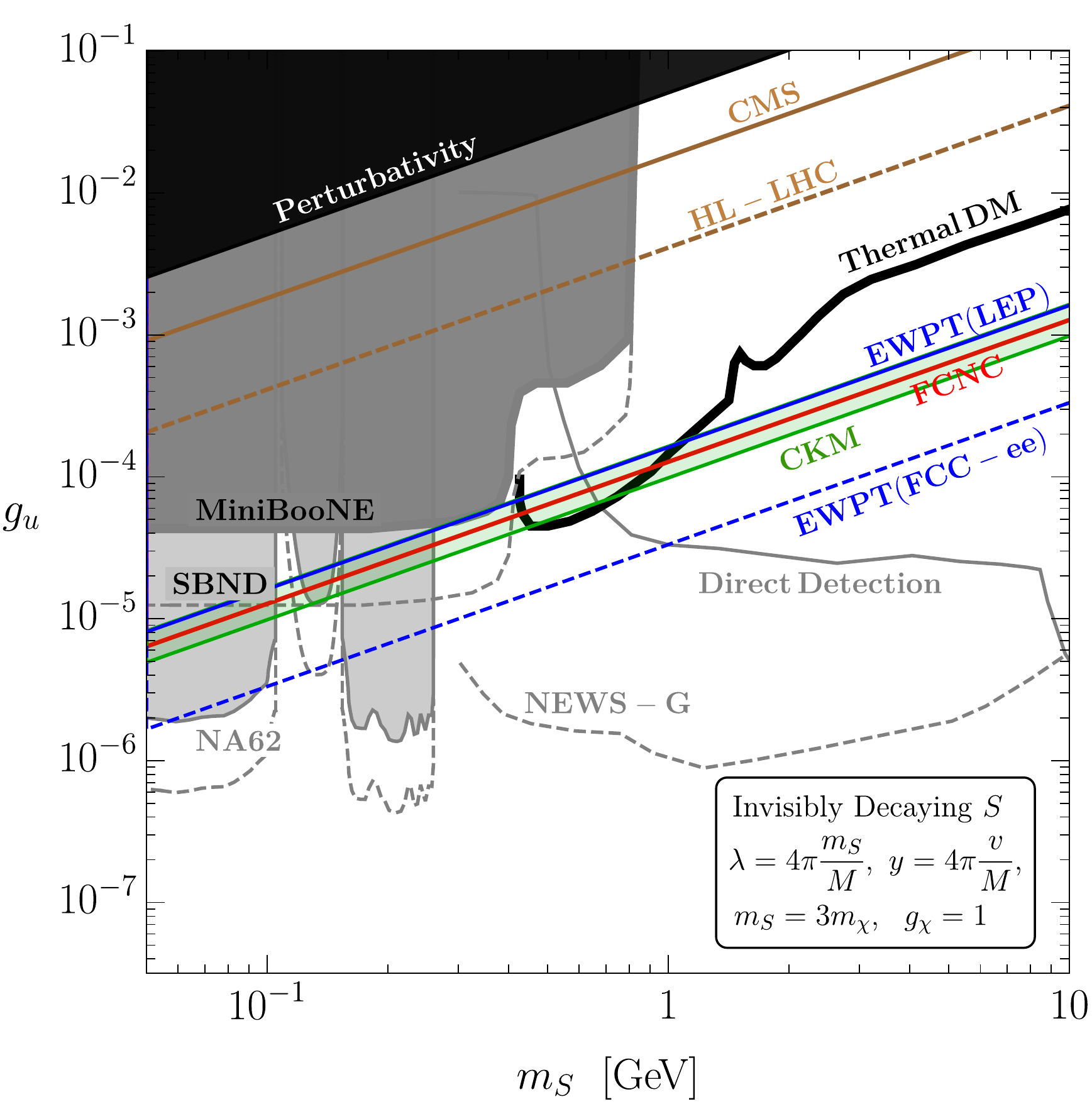}
\caption{\label{fig:VLQlowenplots}
The up-specific scalar EFT parameter space shown in the $m_S-g_u$ plane. The left panels assume the scalar decays visibly to hadrons, while the right panels assume the scalar decays invisibly to dark matter with $g_{\chi} = 1$ and $m_S = 3 m_\chi$. In the top panels,  $y$ is varied while the VLQ mass is fixed to $M = 2$ TeV and $\lambda$ is chosen to saturate the naturalness condition (\ref{eq:lambda-bound}). In the bottom panels, $M$ is varied while both $\lambda$ and $y$ are chosen to saturate their naturalness bounds (\ref{eq:lambda-bound},\ref{eq:y-bound}). In all panels we show several model-independent constraints from Ref.~\cite{Batell:2018fqo} on the EFT parameter space, which depend only on $g_u$ and $m_S$. In addition, constraints from the VLQ model are shown under the stated assumptions for each plot. Further details are given in the main text.
}
\end{figure}
%%%%%%%%%%%%%%%%%%%%

Here we summarize the current bounds and future expected sensitivities in the VLQ completion of the light up-specfic scalar. As argued above in Sec.~\ref{sec:VLQ-naturalness}, for a light scalar satisfying naturalness conditions (\ref{eq:lambda-bound},\ref{eq:y-bound}), we typically expect $\lambda \ll y$. In this situation, the strongest constraints on the UV completion pertain to the coupling $y$ and the VLQ mass $m_{U'} \simeq M$. These limits are summarized in Fig.~\ref{fig:VLQhighenplots}, where we show the constraints from FCNCs in the neutral kaon system, CKM-top row unitarity, $Z$ boson hadronic width, and direct searches at the LHC. The LHC constraint relies on QCD production and thus is not sensitive to the precise value of the coupling $y$, again provided that $\lambda \ll y$. The indirect bounds from FCNCs, CKM unitarity, and EWPT all probe similar regions of parameter space and are generally more stringent for lighter VLQs. As discussed in Sec.~\ref{sec:VLQ-CKM}, the model can explain the $\sim 3\sigma$ discrepancy in the CKM top row unitarity determination for couplings $y\sim 0.1 - 1$ in the mass range 600 GeV $\lesssim M \lesssim  $ 5 TeV, as indicated by the green band in Fig.~\ref{fig:VLQhighenplots}. This region can be probed further at the HL-LHC and definitively tested by a future FCC-ee measurement of $R_\ell$. 

The bounds on the UV completion shown in Fig.~\ref{fig:VLQhighenplots} can also be interpreted within the up-specific scalar EFT mass -- coupling parameter space. Several such interpretations are presented in Fig.~\ref{fig:VLQlowenplots}, which shows a variety of constraints in the $m_S-g_u$ plane. In particular, we show both the model-independent constraints relying only on $g_u$ and $m_S$ derived previously in Ref.~\cite{Batell:2018fqo} (see the next paragraph for details), along with the constraints depending on the VLQ UV completion. The left panels assume the scalar decays visibly to hadrons, while the right panels assume the scalar decays invisibly to dark matter with $g_{\chi} = 1$ and $m_S = 3 m_\chi$. In the top panels,  $y$ is varied while the VLQ mass is fixed to $M = 2$ TeV and $\lambda$ is chosen to saturate the naturalness condition (\ref{eq:lambda-bound}). Therefore, the top panels always satisfy the direct constraints from the LHC on VLQs, but can only satisfy the naturalness conditions if the scalar is sufficiently weakly coupled. In contrast, in the bottom panels $M$ is varied while both $y$ and $\lambda$ are chosen to saturate their naturalness bounds (\ref{eq:lambda-bound},\ref{eq:y-bound}). With these assumptions, all parameters shown in the the bottom panels are natural, but LHC VLQ searches rule out low mass, strongly coupled scalars. Regions shown in black correspond to nonpertubative values of the coupling, $y>4\pi$. One observes that bounds from the VLQ completion uniquely probe certain regions of the light scalar parameter space. These bounds are therefore highly complementary to those obtained in the EFT analysis ~\cite{Batell:2018fqo}. 

Finally, we provide a brief summary of the constraints on the low energy scalar EFT appearing in Fig.~\ref{fig:VLQlowenplots}; see Ref.~\cite{Batell:2018fqo} for more details. We first discuss the case of visible scalar decays (left panels). Scalars lighter than the di-pion threshold will decay radiatively to a pair of photons and tend to be long-lived for natural values of the coupling. This low mass region is tightly constrained by fixed target experiments (CHARM~\cite{Bergsma:1985qz}), rare pion decays (MAMI~\cite{Nefkens:2014zlt}), Big Bang nucleosynthesis, and supernova data. For masses $m_S > 2 m_\pi$, there are constraints from rare $\eta$ (KLOE~\cite{Anastasi:2016cdz}) and $\eta'$ (BESII~\cite{Ablikim:2016frj}) decay searches, while future $\eta$ decay searches at REDTOP~\cite{Gatto:2016rae,Gan:2020aco} will test a currently viable and natural region of parameter space. In addition, searches for long-lived scalars at FASER/FASER2~\cite{Kling:2021fwx,Feng:2017uoz} and the proposed SHiP experiment~\cite{Alekhin:2015byh} can probe very feeble couplings. Finally, if there is a new ${\cal O}(1)$ CP-violating phase in the theory, the neutron EDM constraint discussed in Section~\ref{sec:VLQ-CPviolation} provides the 
 strongest bound today for $m_S > 2 m_\pi$.

For the case of invisible scalar decays to dark matter particles (right panels), searches for the rare kaon decay, $K\rightarrow \pi S, S\rightarrow$ invisible, at NA62~\cite{NA62:2020pwi,NA62:2021zjw} provide the best constraint at low masses, while substantial improvements are anticipated in the near future with the full NA62 dataset. The MiniBooNE beam dump dark matter search and a future beam dump run at SBND can provide powerful tests in the several hundred MeV mass range~\cite{Aguilar-Arevalo:2017mqx,Aguilar-Arevalo:2018wea,vandewater}. At larger masses of order GeV and above, direct detection experiments such as CRESST-III~\cite{CRESST:2019axx}, DAMIC~\cite{DAMIC:2020cut},
XENON1T~\cite{XENON:2019gfn},
PandaX~\cite{PandaX-II:2017hlx}, and in the future NEWS-G~\cite{Arnaud:2017bjh,Battaglieri:2017aum}, will provide the leading constraints in this simple hadrophilic dark matter model. Also shown in the right panels of Fig.~\ref{fig:VLQlowenplots} are the parameters leading to the correct dark matter thermal relic abundance through freezeout of dark matter annihilation to hadrons. We observe that low-energy EFT probes as well as a number of 
measurements unique to the VLQ completion can provide complementary tests of the cosmologically motivated region of parameter space.

%%%%%%%%%%%%%%%%%%%%%%%%%%%%%%%%%%%%%%%%
%%%%%%%%%%%%%%%%%%%%%%%%%%%%%%%%%%%%%%%%
\section{Scalar Doublet Completion}
\label{sec:SD-completion}

In this section we investigate a second renormalizable completion of the flavor-specific EFT involving an additional scalar electroweak doublet.
After presenting the model, we discuss the expected radiative contributions to the couplings and the ensuing naturalness criteria. We then study the minimization of the potential, the passage to the physical basis, and the decays of the new scalar doublet states. A study of the phenomenology follows, including the predictions and constraints from electroweak precision tests, FCNCs, CP violation, and searches for the the new scalars at the LHC. Finally, we conclude this section with a summary of these bounds along with several interpretations in the low-energy scalar EFT parameter space.

%%%%%%%%%%%%%%%%%%%%%%%%%%%%%%%%%%%%%%%%
\subsection{Model}

We consider a model with a singlet scalar $S$ and a heavy scalar mediator with the same quantum numbers as the Higgs, $H' \sim ({\bf 1}, {\bf 2}, \tfrac{1}{2})$.
The minimal Lagrangian is given by
\begin{align}
\Lagr_{\rm sd} = \Lagr_{\rm SM} &+ \frac{1}{2} \partial_\mu S \partial^\mu S - \frac{1}{2} m_S^2 S^2 + (D_\mu H')^\dagger D^\mu H' - M^2 H'^\dagger H' 
\notag \\
& -\bigl[ {y'}_i^j \, \overline Q_L^{\, i}
\,u_{R \, j}\,H'_{c}+\kappa \, M\,S\,H^{\dagger}H^{'}
+\text{h.c.}\bigr] + \text{quartic scalar couplings},
\label{eq:L-HH}
\end{align}
where $i = 1,2,3$ is a generation index and $M$ is the mass of the scalar doublet. To render $\kappa$ dimensionless, the scalar triple coupling has been re-scaled with $M$. Integrating out the scalar doublet at tree-level, we obtain the leading effective interactions at dimension 4 and 5: 
\begin{equation}
{\cal L} \supset |\kappa|^2 S^2 |H|^2+\frac{\kappa \, {y'}_i^j}{M} S \,\overline Q_L^i \, u_{R \,j} \, H_c +{\rm h.c.}
\label{eq:L-EFT-doublet}
\end{equation}
The first term in (\ref{eq:L-EFT-doublet}) is the Higgs portal operator, which we will return to in the next subsection when we discuss the scalar potential. The second term in (\ref{eq:L-EFT-doublet}) gives rise to the scalar-quark coupling of interest. 
Thus, we can identify $M$ with the new scale and $(c_S)_i^j = - \kappa\, {y'}_i^j$, respectively, in the effective operator~(\ref{eq:bsmlag}). In the flavor basis in which the SM up quark Yukawa couplings are diagonal, the up-specific hypothesis corresponds to ${y'}_i^j \propto \delta_{i1} \delta^{j1}$. As with the VLQ model, we provide a description of the flavor hypothesis employed in the scalar doublet completion leading to the starting Lagrangian~(\ref{eq:L-HH}) in Appendix~\ref{app:SD-flavor-hypothesis}.

Similar to the VLQ model in Eq.~(\ref{eq:L-VLQ-UR}), the Lagrangian in Eq.~\eqref{eq:L-HH} could be extended by additional renormalizable scalar potential terms involving $S$, $H$ and/or $H'$. In the absence of fine-tuning, small but non-zero coefficients of these terms are induced radiatively, as will be discussed below. However, we will assume that they do not receive any tree-level contribution that is parametrically larger than these loop effects.

%%%%%%%%%%%%%%%%%%%%%%%%%%%%%%%%%%%%%%%%
\subsection{Naturalness considerations}
\label{scalarnat}

We now consider the implications of naturalness on the scalar potential, following the same philosophy and approach used for the VLQ model; see Sec.~\ref{sec:VLQ-naturalness}. Our aim is to estimate the expected radiative sizes of the various scalar interactions generated by the couplings of $S$ and $H$ to the heavy scalar doublet $H'$ in \eqref{eq:L-HH}. As in Sec.~\ref{sec:VLQ-naturalness}, the size of the loop corrections are estimated by including factors  of $(16\pi^2)^{-1}$ for each loop and counting the pertinent coupling and scale factors, the latter of which are taken to be $M$.

\begin{figure}
    \includegraphics[width=0.2\textwidth]{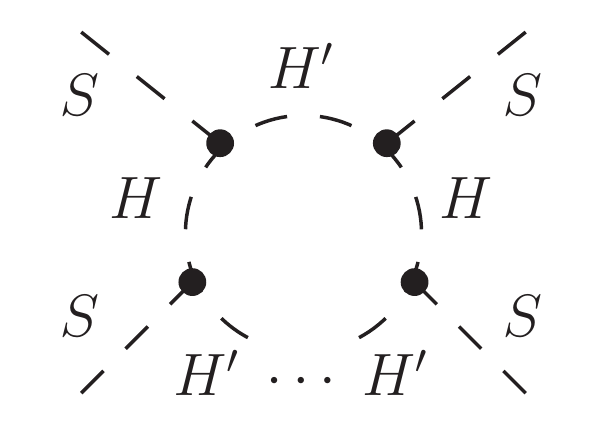} \hfill
    \includegraphics[width=0.2\textwidth]{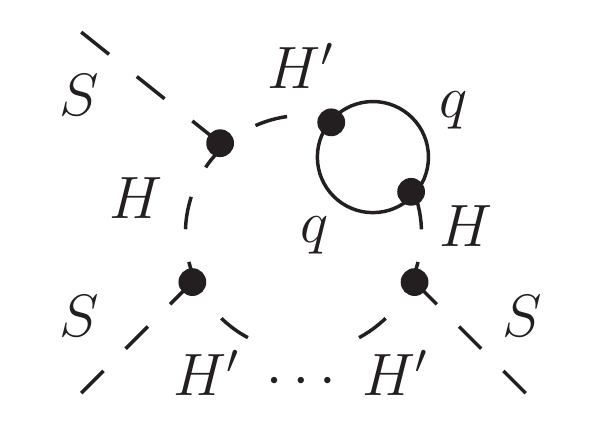} \hfill
    \includegraphics[width=0.22\textwidth]{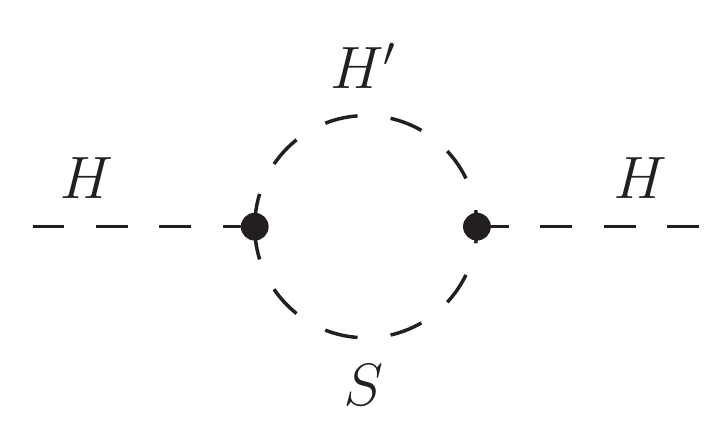} \hfill
    \raisebox{-1ex}{\includegraphics[width=0.2\textwidth]{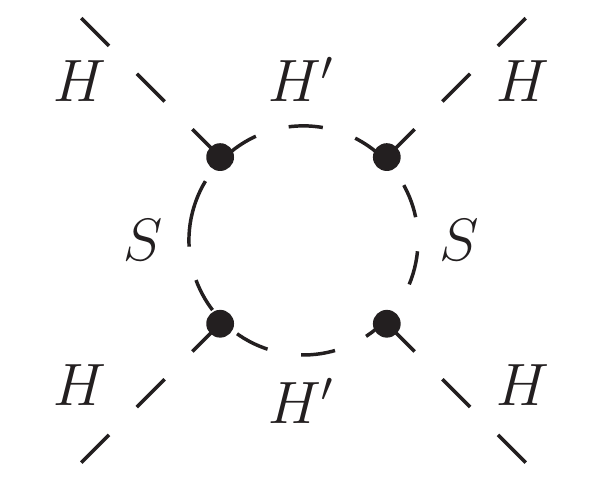}} \\
    \makebox[0.2\textwidth]{\centering (a)} \hfill \makebox[0.2\textwidth]{\centering (b)}\hfill \makebox[0.22\textwidth]{\centering (c)} \hfill \makebox[0.2\textwidth]{\centering (d)}
    \caption{Loop-induced contributions to scalar self-couplings in the scalar doublet UV completion.}
    \label{fig:scaln}
\end{figure}

For interactions of even or odd numbers of the scalar $S$ one thus finds \begin{align}
&\delta_{S^{2k}} \sim \frac{|\kappa|^{2k}}{16\pi^2}M^{4-2k}, 
&
&\delta_{S^{2k+1}} \sim \frac{|\kappa|^{2k}\,\text{Re}\{\kappa
\Tr(y'y_u^\dagger)\}}{(16\pi^2)^2}M^{3-2k}, \label{eq:Sn}
\end{align}
from the one- and two-loop diagrams in Fig.~\ref{fig:scaln}~(a) and (b), respectively. The case $k=1$ corresponds to a correction to the mass parameter, $m_S^2$, given by $\delta m_S^2 \sim |\kappa|^2M^2/16\pi^2$. Requiring $\delta m_S^2$ to be less than the physical mass $m_S^2$ leads to the bound
\be
|\kappa| \lesssim 4\pi \frac{m_S}{M}  \simeq  (6 \times 10^{-3})   \left(  \frac{m_S}{1\, \rm GeV}  \right)  \left(  \frac{2 \, \rm TeV}{M}  \right) . \label{eq:kappabound1}
\ee
This can be compared to the tree-level contribution from the Higgs portal operator, which arises from integrating out the heavy scalar doublet, Eq.~(\ref{eq:L-EFT-doublet}). After electroweak symmetry breaking, this gives a correction to the scalar mass, $\delta m_S^2 \sim |\kappa|^2 v^2$, leading to the naturalness condition
\be
|\kappa| \lesssim \frac{m_S}{v} 
 \simeq  (4 \times 10^{-3})   \left(  \frac{m_S}{1\, \rm GeV}  \right). 
 \label{eq:kappabound2}
\ee
This condition is stronger than (\ref{eq:kappabound1}) unless $M \gtrsim 4 \pi v$. We note that there is no analogous one-loop naturalness condition on the coupling $y'$. However, at two loops there is a contribution to the $S$ mass depending on both $\kappa$ and $y'$, which precisely corresponds to the two-loop correction in the EFT that was mentioned in Eq.~(\ref{eq:natural-criterion}).

Similarly to the singlet scalar, the one-loop diagrams in Fig.~\ref{fig:scaln}~(c) and (d) lead to corrections to the SM Higgs mass and self-coupling,
\begin{align}
&\delta \mu^2 \sim \frac{|\kappa|^2}{16\pi^2}M^2,
&
&\delta \lambda \sim \frac{|\kappa|^4}{16\pi^2}.
\end{align}
Demanding that $\delta \mu^2 \lesssim \mu^2 = m_h^2/2$ leads to the bound $\kappa \lesssim 2^{3/2}\pi \,m_h/M$, which is a weaker bound than \eqref{eq:kappabound1} for $m_S \sim {\cal O}$(GeV).
%------------------------------
%------------------------------

In addition, \eqref{eq:Sn} generates a number of scalar self-interaction terms that were not present in the original Lagrangian \eqref{eq:L-HH}: 
\begin{align}
\label{eq:S-terms}
{\cal L} &\supset -\delta_S S - a_3 S^3 - a_4 S^4, \\
\delta_S &\sim \frac{\text{Re}\{\kappa \Tr(y'y_u^\dagger)\}}{(16\pi^2)^2}M^3, \label{eq:S-delta}\\
a_3 &= \frac{|\kappa|^2 \,\text{Re}\{\kappa \Tr(y'y_u^\dagger)\}}{(16\pi^2)^2}M, \qquad a_4 = \frac{|\kappa|^4}{16\pi^2}.  \label{eq:S-a3-a4}
\end{align}
The presence of the tadpole term $\delta_S$ causes $S$ to develop a vev, $v_S$. The couplings $a_3$ and $a_4$ also have an influence on the value of $v_S$, but it is subdominant given the radiative estimates in Eq.~(\ref{eq:S-a3-a4}) for values of $\kappa$ that satisfy the naturalness bound in Eqs.~(\ref{eq:kappabound1},\ref{eq:kappabound2}).

In a similar fashion, there are radiatively generated $S|H|^2$, $S^2|H|^2$, $S|H'|^2$, $S^2|H'|^2$ and $|H'|^4$ terms, which can be neglected to first approximation in phenomenological applications. More relevant is the loop-induced mixing mass term
\begin{align}
{\cal L} &\supset \delta \mu'^2\,H'^\dagger H + \text{h.c.},
&
\delta \mu'^2 &\sim \frac{\Tr(y'y_u^\dagger)}{16\pi^2}M^2 .
\end{align}

%%%%%%%%%%%%%%%%%%%%%%%%%%%%%%%%%%%%%%%%
\subsection{Mixing and mass eigenstates}

\subsubsection{Scalar potential}

Including the leading radiatively induced tadpole and mass terms from the previous subsection, but neglecting the loop corrections to 3- and 4-point interactions, the scalar potential takes the form 
\begin{equation}
V\supset -\mu^{2}(H^{\dagger}H)+\lambda(H^{\dagger}H)^{2}+ \delta_{S}S+\frac{m_S^{2}}{2}S^{2}+M^{2}(H'^{\dagger}H')
+ \left[-\delta\mu'^{2}H'^{\dagger}H + \kappa M(H'^{\dagger}H)S+\text{h.c.}\right] 
\end{equation}
In general, the neutral components of all three scalar fields produce vevs, for which we introduce the following notation:
$\langle S \rangle=v_{S}$, $\langle H\rangle=(0,v_{0}/\sqrt{2})^\top$,
$\langle H'\rangle=(0,v'/\sqrt{2})^\top$. Minimizing the scalar potential, we can solve for the vevs for $S$ and $H'$ to get 
\begin{align}
& v'=\frac{v_{0}\left(\delta_{S}\,\kappa M+m_S^{2}\,\delta\mu'\,^{2}\right)}{M^{2}\left(m_S^{2}-\kappa^{2}v_{0}^{2}\right)},
& &v_{S}=-\frac{\delta_{S}M+\kappa\,\ensuremath{\delta\mu'}\,^{2}v_{0}^{2}}{M\left(m_S^{2}-\kappa^{2}v_0^{2}\right)}.
\end{align}
One can reduce the number of independent parameters by using the radiative estimates $\kappa \lesssim 4\pi m_S/M$, $\delta_S \sim \frac{M^3\kappa}{(16\pi^2)^2} \Tr(y'y_u)$, $\delta\mu'^2 \sim \frac{M^2}{16\pi^2}
\Tr(y'y_u)$ from section \ref{scalarnat}. The outcome depends on the relative sign between $\delta_S$ and $\delta\mu'^2$ (which in general is unknown since both terms can receive additional tree-level contributions). However, in the limit of large $M$ the expressions simplify to 
\begin{align}
\biggl|\frac{v'}{v_0}\biggr| &\sim\frac{y'\, y_{u}}{16\pi^{2}} \sim 10^{-7}, &
\biggl|\frac{v_S}{v_0}\biggr| &\sim \frac{  y'\, y_{u} M^2  }{64\pi^3 m_S v_0}
 \sim 10^{-4} 
\label{eq:vevapprox}
\end{align}
where we have specialized to the up-specific scenario and assumed $y' \sim {\cal O}(1)$, $m_S \sim {\cal O}$(GeV) and $M > 1$~TeV for the numerical estimates.

Similar to the 2HDM, it is useful to rotate the doublets to the ``Higgs
basis'', where only one of the doublets develops a vev, while the singlet remains unchanged, viz. : 
\be
\begin{pmatrix}
\hat{H}\\
\hat{H'}
\end{pmatrix}=\begin{pmatrix}
\cos\beta & \sin\beta \\
-\sin\beta & \cos\beta
\end{pmatrix}\begin{pmatrix}
H\\
H'
\end{pmatrix},
\ee
where $\tan\beta\equiv v'/v_{0}$. The fields can be decomposed according to
\begin{align}
&\hat{H} = \begin{pmatrix} G^+ \\ \frac{1}{\sqrt{2}}(v+\phi_1 + iG^0) \end{pmatrix},
&&\hat{H}' = \begin{pmatrix} H^+ \\ \frac{1}{\sqrt{2}}(\phi_2 + iA^0) \end{pmatrix},
&& S = v_S + \phi_3,
\label{eq:hcomp}
\end{align}
where $v = 246$ GeV. 
The CP-even scalar fields $\phi_1, \phi_2, \phi_3$ will mix with each other. Diagonalizing their $3\times3$ mass matrix ${\cal M}^2_\phi$ leads to three mass eigenstates $h,h',s$, 
\be
R^{T}\mathcal{M}^{2}_\phi R=\text{diag}\{m_{h}^{2},m_{h'}^{2},m_{s}^{2}\},
\ee
where $h$ corresponds to the SM-like Higgs boson discovered at the LHC. For $\tan\beta \ll 1$, we can approximately write 
\be
\mathcal{M}^{2}_\phi \simeq \begin{pmatrix}
 2 \lambda v^2 & -2 \lambda v^2 \tan\beta 
& 2 \kappa  M v \tan\beta  \\
 -2 \lambda v^2 \tan\beta
 & M^2 & \kappa  M v  \\
 2 \kappa  M v \tan\beta & \kappa  M v  & m_S^2 \\
\end{pmatrix}
\ee
Since the off diagonal terms are small, the rotation matrix takes the approximate form 
\begin{align}
&R\simeq \begin{pmatrix}
1 & \theta_{12} & \theta_{13}\\
-\theta_{12} & 1 & \theta_{23}\\
-\theta_{13} & -\theta_{23} & 1
\end{pmatrix},
&&\text{with}
&&\begin{array}{l}
\theta_{12} \simeq 
-2\lambda v^2 \tan\beta/M^2, \\
\theta_{13} \simeq
\kappa M \tan\beta / \lambda v, \\
\theta_{23} \simeq 
- \kappa v/M,
\end{array}
\end{align}
where we have kept the leading contributions to the mixing angles in the limit $m_S^2 \ll \lambda v^2 \ll M^2 $ and $\tan\beta \ll 1$. Similarly, one finds that the CP-even scalar masses are approximately given by 
\be
m_{h}^2 \simeq 2 \lambda v^2, \qquad m_{h'}^2 \simeq M^2, \qquad m_{s}^2 \simeq m_S^2-\kappa^2v^2.
\label{eq:m-CP-even}
\ee
Note that the second contribution to the light singlet squared mass eigenstate comes from the Higgs portal operator in Eq.~(\ref{eq:L-EFT-doublet}). We will always impose $|\kappa|<m_S/v$ such that $m_s^2 >0$ in what follows. The masses of $A^0$ and $H^\pm$ are given by
\be
m_{A^0,H^\pm}^2 = \frac{M^2}{\cos^2\beta}  \approx M^2.
\label{eq:mA0-mHPM}
\ee

\subsubsection{Scalar decays}

It is straightforward to work out the interaction Lagrangian in the mass basis. However, since the expectation is that the mixing between the scalar doublets is small, i.e., $\tan\beta \ll 1$, many of the phenomenological consequences can be extracted directly from our starting Lagrangian, Eq.~(\ref{eq:L-HH}). Here we consider the decays of the heavy scalar. While in general 2HDMs gauge interactions often mediate decays of a heavy scalar doublet component into a lighter doublet component and an electroweak boson $(W,Z,h)$, such two-body decays are typically kinematically forbidden in our scenario due to the approximate mass degeneracy of the doublet components (see Eqs.~(\ref{eq:m-CP-even},\ref{eq:mA0-mHPM})). The leading decays of the scalar doublet then arise from the new couplings $y'$ and $\kappa$ in Eq.~(\ref{eq:L-HH}). These lead to the partial widths
\begin{align}
\label{eq:SD-decay-quarks}
\Gamma(h'\rightarrow u \bar u) = \Gamma(A^0\rightarrow u \bar u) = \Gamma(H^+\rightarrow u \bar d)& \simeq \frac{3 y'^2 M}{16 \pi}, \\
\label{eq:SD-decay-bosons}
\Gamma(h'\rightarrow s h ) = \Gamma(A^0\rightarrow s Z) = \Gamma(H^+\rightarrow s W^+) &\simeq \frac{\kappa^2 M}{16 \pi}.
\end{align}
These expressions are valid in the limit $\tan\beta \ll 1$ and $M \gg v$. In natural regions of parameter space, we expect that $\kappa$ satisfies the conditions (\ref{eq:kappabound1},\ref{eq:kappabound2}) and is typically much smaller than $y'$, which is not subject to any analogous naturalness condition. In this case, the decays of the doublet to first-generation quarks will dominate. This will lead to a dijet resonance signature at the LHC, which we will discuss in more detail below. 

For completeness, it should be noted that other decays are possible due to mixing of the scalar doublets. In particular, there can be decays of heavy scalar doublet components into pairs of lighter electroweak, Higgs, and singlet bosons. The corresponding partial widths scale as $\tan^2\beta$ and are thus expected to be highly suppressed in natural regions of parameter space. As for the light singlet scalar $s$, it will predominantly decay visibly to pairs of up quarks if there are no lighter hidden sector states. Alternatively, if the scalar couples strongly to light dark matter, it may decay via $s\rightarrow \chi \overline \chi$. See also the discussion in Sec.~\ref{sec:VLQ-mixing}.

%%%%%%%%%%%%%%%%%%%%%%%%%%%%%%%%%%%%%%%%
\subsection{Electroweak precision bounds}

%------------------------------
\begin{figure}
    \includegraphics[width=0.72\textwidth]{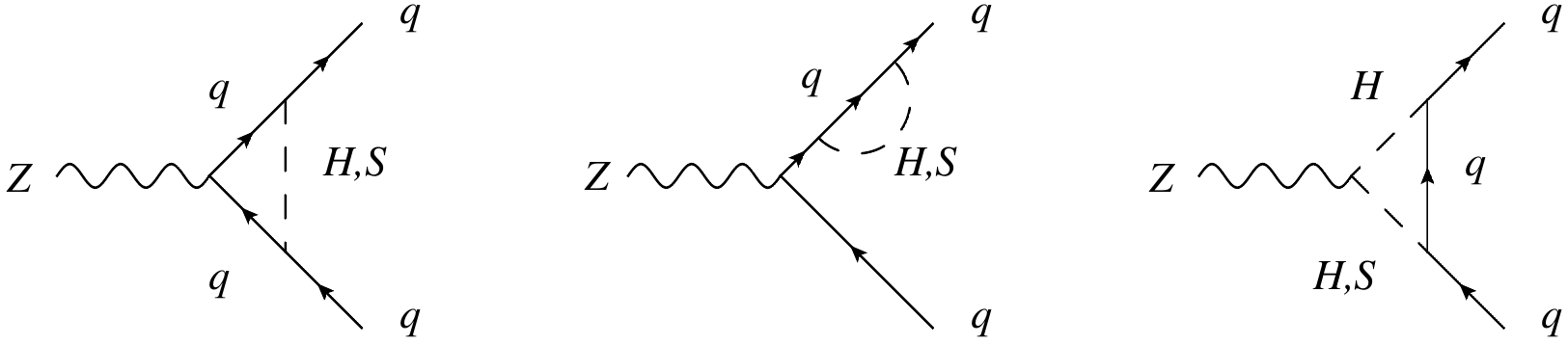}\\
    \makebox[0.26\textwidth]{\centering (a)}  \makebox[0.26\textwidth]{\centering (b)}  \makebox[0.26\textwidth]{\centering (c)}
    \caption{Loop diagrams contributing to the hadronic $Z$ width in the scalar doublet UV completion.}
    \label{fig:ewpS}
\end{figure}
%------------------------------
Similar to the VLQ model, the scalar doublet model modifies the partial width of the $Z$ boson to hadrons.  The leading correction in given by the loop diagrams in Fig.~\ref{fig:ewpS}.
In the limit $M \equiv M_{H'} \gg v \gg m_S$, these yield the following shifts to the $Z$ couplings:
\begin{align}
\label{eq:ewpSuR} 
&\delta g_{uR} \approx g_{uR}^{\rm SM}
\begin{aligned}[t] &\biggl\{ 
\frac{y'^2\,m_Z^2}{48\pi^2\,M^2}
 \biggl[\frac{5}{6} - \ln\Bigl(-\frac{m_Z^2+i\epsilon}{M^2}\Bigr)\biggr] \\
& +\frac{y'^2\kappa^2 v^2}{128\pi^2 M^2}
 \biggl[-\frac{1}{2}-\frac{9}{8s_W^2} + \Bigl(1-\frac{3}{4s_W^2}\Bigr)
 \ln\Bigl(-\frac{m_Z^2+i\epsilon}{M^2}\Bigr) + \ln\frac{M^2}{m_S^2} \biggr]
 \biggr\}, \end{aligned}
 \displaybreak[0]  \\
\label{eq:ewpSuL} 
&\delta g_{uL} \approx g_{uL}^{\rm SM}
\begin{aligned}[t] &\biggl\{  
\frac{y'^2\,m_Z^2}{(18-24s_W^2)\pi^2\,M^2}
 \biggl[\frac{1}{8}-\frac{s_W^2}{3} + s_W^2
  \ln\Bigl(-\frac{m_Z^2+i\epsilon}{M^2}\Bigr)\biggr] \\
 & +\frac{y'^2\kappa^2 v^2}{128\pi^2 M^2} 
 \biggl[-\frac{6-2s_W^2}{3-4s_W^2} - \frac{4s_W^2}{3-4s_W^2}
 \ln\Bigl(-\frac{m_Z^2+i\epsilon}{M^2}\Bigr) + \ln\frac{M^2}{m_S^2} \biggr]
 \biggr\}, \end{aligned} 
 \displaybreak[0] \\
&\delta g_{dL} \approx g_{dL}^{\rm SM}\, \frac{y'^2\,m_Z^2}{(18-12s_W^2)\pi^2\,M^2}
 \biggl[\frac{1}{8}+\frac{s_W^2}{12} - s_W^2
  \ln\Bigl(-\frac{m_Z^2+i\epsilon}{M^2}\Bigr)\biggr].
  \label{eq:ewpSdL} 
\end{align}
The second lines in \eqref{eq:ewpSuR} and \eqref{eq:ewpSuL} are additionally suppressed by $\kappa^2$ but they are enhanced by the logarithm $\ln M^2/m_S^2$. 

Plugging Eqs.~(\ref{eq:ewpSuR},\ref{eq:ewpSuL},\ref{eq:ewpSdL}) into Eq.~(\ref{eq:dRl}), we obtain the correction to the $Z$ boson hadronic-to-leptonic branching ratio $R_\ell$.  For $M = 1\,\text{TeV}$, $m_S = 1\,\text{GeV}$ and $y' = \kappa = \sqrt{4\pi}$ one finds that $R_\ell$ is shifted by 0.83, which is excluded by current data, $R_\ell^{\rm exp} - R_\ell^{\rm SM} = 0.034\pm 0.025$ \cite{Tanabashi:2018oca}.
For $y' = \kappa = 1$, the shift is instead $5.5 \times 10^{-3}$, which is
currently not excluded and can be probed only marginally by FCC-ee, with an expected $1\sigma$ precision of $\delta R_\ell^{\rm exp.} = 0.001$ \cite{Blondel:2018mad}.

%%%%%%%%%%%%%%%%%%%%%%%%%%%%%%%%%%%%%%%%
\subsection{FCNCs}

Similar to the FCNC we discussed in the VLQ section, there is a one loop box diagram resulting from $H'$ and up quark exchange, which leads to an effective operator with four $Q_L$ fields given by
\begin{equation}
{\cal L} \supset - \frac{(y' y'^{ \dag })_i^j(y' y'^{ \dag })_k^l}{128 \pi^2 M^2} [ \overline Q^{\, i} \gamma^\mu P_L \, Q_{\, j}][ \overline Q^{\, k} \gamma^\mu P_L \, Q_{\, l}].
\end{equation}
With the up-specfic hypothesis and moving to the physical basis, we obtain a contribution to neutral Kaon mixing, described by the operator in Eq.~(\ref{eq:kaon}) with Wilson coefficient
$C^{ds} =  - (y')^4 |V_{ud}^* V_{us}|^2/(128 \pi^2 M^2$). Applying the bound ${\rm Re}[C^{ds}] \lesssim (10^3 \, {\rm TeV})^{-2}$~\cite{Bona:2007vi}, we obtain the constraint
\begin{equation}
y' \lesssim 0.6 \left(\frac{M}{2 \, {\rm TeV}}\right)^{1/2},
\end{equation}
similar to Eq.~(\ref{eq:vlqfcnc}) for the VLQ model.

%%%%%%%%%%%%%%%%%%%%%%%%%%%%%%%%%%%%%%%%
\subsection{CP violation}

In the scalar doublet completion, the basis independent CP-violating phase is 
\begin{align}
    \phi_{\rm CP}&=\arg\left(y_u {y^\prime}^\ast\kappa \right)
\end{align}
%%%%%%%%%%%%%%%%%%%%
Separate rephasings of $u_{L,R}$ and $H^\prime$ leave this quantity invariant. If $\phi_{\rm CP}$ is nonvanishing, a nonzero neutron EDM will develop. This occurs in much the same way as in the VLQ completion, namely through a CP-violating four up quark operator mediated by $S$ exchange. This operator is defined in Eq.~(\ref{eq:CPodd4up}). In this model, the corresponding Wilson coefficient is 
\begin{equation}
\label{eq:CPodd4upWilson-SD}
C'_u  \simeq - \frac{y'^2 \kappa^2 v^2}{4M^2 m_S^2 }\sin{2\phi_{\rm CP}}.
\end{equation}
Using Eqs.~(\ref{eq:neutronEDM},\ref{eq:CPodd4upWilson-SD}) we can express this as a limit on the effective coupling of the scalar to up quarks ($g_u \simeq y' \kappa v /\sqrt{2} M$). We obtain the same bound as in the VLQ model given in 
Eq.~(\ref{eq:neutronEDM4up}).

%%%%%%%%%%%%%%%%%%%%%%%%%%%%%%%%%%%%%%%%
\subsection{Collider phenomenology}

We next discuss signatures of the heavy scalar doublet at the LHC. Motivated by the naturalness conditions (\ref{eq:kappabound1},\ref{eq:kappabound2}), we typically expect $\kappa\ll y$, in which case the scalar doublet will decay to first-generation quarks through the $y'$ coupling; see Eqs.~(\ref{eq:SD-decay-quarks}) for the partial decay widths. This makes it challenging to probe the scalar doublet through its electroweak pair production process at the LHC, given the low production rate and large QCD backgrounds. On the other hand, if $y'$ is large enough the heavy scalar doublet can be produced singly in quark-antiquark annihilation and decays into a di-jet final state. Since all physical eigenstates of the heavy doublet have masses that are very close to each other, $m_{h'}\approx m_{A^0,H^\pm} \approx M$, and they all decay dominantly into quarks, they would manifest as a single narrow\footnote{Here ``narrow'' means that the physical decay width of all heavy scalars is smaller than the experimental resolution.} di-jet resonance. The influence of the mixing angle $\beta$ is very small and can be safely neglected in this context.

Both ATLAS and CMS have conducted searches for di-jet resonances at $\sqrt{s}=13$~TeV and presented bounds in terms of several representative models \cite{Aad:2019hjw,CMS:2018mgb,Sirunyan:2019vgj}. We use the published bounds for hadro-philic $Z'$ models to derive corresponding limits for the heavy scalar doublet. For this purpose, we have computed fiducial cross-sections for both the $Z'$ model and the scalar doublet model with CalcHEP 3.4.6 \cite{Belyaev:2012qa}, for a grid of different resonance masses ranging from 100~GeV to 7~TeV. Since both cases are $q\bar{q}$ initiated, one may expect that the K-factor from QCD corrections is similar for both models and cancels when taking the ratio of the cross-sections.  We then used these cross-section ratios to re-scale the coupling limits for the $Z'$ model reported in Refs.~\cite{Aad:2019hjw,CMS:2018mgb,Sirunyan:2019vgj}. For the low-mass region, below 500~GeV, a boosted di-jet search by CMS can be utilized \cite{Sirunyan:2019vxa}. Furthermore, the HL-LHC will be able to extend the reach to di-jet resonances, particularly in the high mass region. We have translated one such HL-LHC projection from ATLAS to the scalar doublet model~\cite{ATL-PHYS-PUB-2015-004}. This translation depends on the K-factor for $pp\to H'$, which is currently unknown. For simplicity, we have used $K=1$, which is supported by the fact that the closely related Drell-Yan (see e.g.\ Ref.~\cite{Bonvini:2010tp}) and scalar diquark production \cite{Han:2009ya} processes have small K-factors of about 1.2. The resulting limits and projections on the Yukawa coupling $y'$, as a function of the mass $M$, are shown in Fig.~\ref{fig:scalarhighenplots}.

Let us also make a few comments about the scenario that $y' < \kappa$. In this case  the scalar doublet decays predominantly to an electroweak or Higgs boson and $s$, see Eq.~(\ref{eq:SD-decay-bosons}). Furthermore, the condition $y' < \kappa$ combined with the naturalness constraints on $\kappa$ suggest that $y'$ is relatively small in this scenario, such that the single production process $q\bar{q} \to H'$ is suppressed. In this case heavy scalar pair production, mediated by electroweak gauge interactions, may be more promising. $H^\pm h'$ and $H^\pm A^0$ production, followed by the decays $H^\pm \to sW^\pm$, $h'\to sh$ and $A^0 \to sZ$, leads to final states with several leptons and/or a $b\bar{b}$ pair. If $s$ decays into light dark matter particles, these signatures are very similar to gaugino pair production processes in the MSSM. Thus we expect that heavy scalar masses $M \lesssim {\cal O}$(TeV) are excluded by $\tilde{\chi}_1^\pm\tilde{\chi}^0_2$ searches at ATLAS and CMS \cite{ATLAS:2021susy,CMS:2021lzg,CMS:2020bfa}, but the details of this bound depend on the different production cross-sections in the MSSM and our scalar doublet model. If instead $s$ decays visibly into hadrons, the signature is very similar to the VLQ searches discussed in section~\ref{sec:VLQ-collider}, with the main difference that the heavy scalar pair production is an electroweak rather than a strong process. As a result, we expect somewhat weaker limits than those reported for VLQs in section~\ref{sec:VLQ-collider}.

Finally, as in the VLQ model, the singlet scalar $s$ can be produced directly at the LHC and show up as either a di-jet resonance if it decays visibly or as a mono-jet if it decays invisibly. 
In both cases the limit on the effective coupling $g_u$ is rather weak. For further details, see the earlier discussion in Sec.~\ref{sec:VLQ-collider}.

\subsection{Summary}

%%%%%%%%%%%%%%%%%%%%
\begin{figure}
\centering
\includegraphics[width=0.7\textwidth]{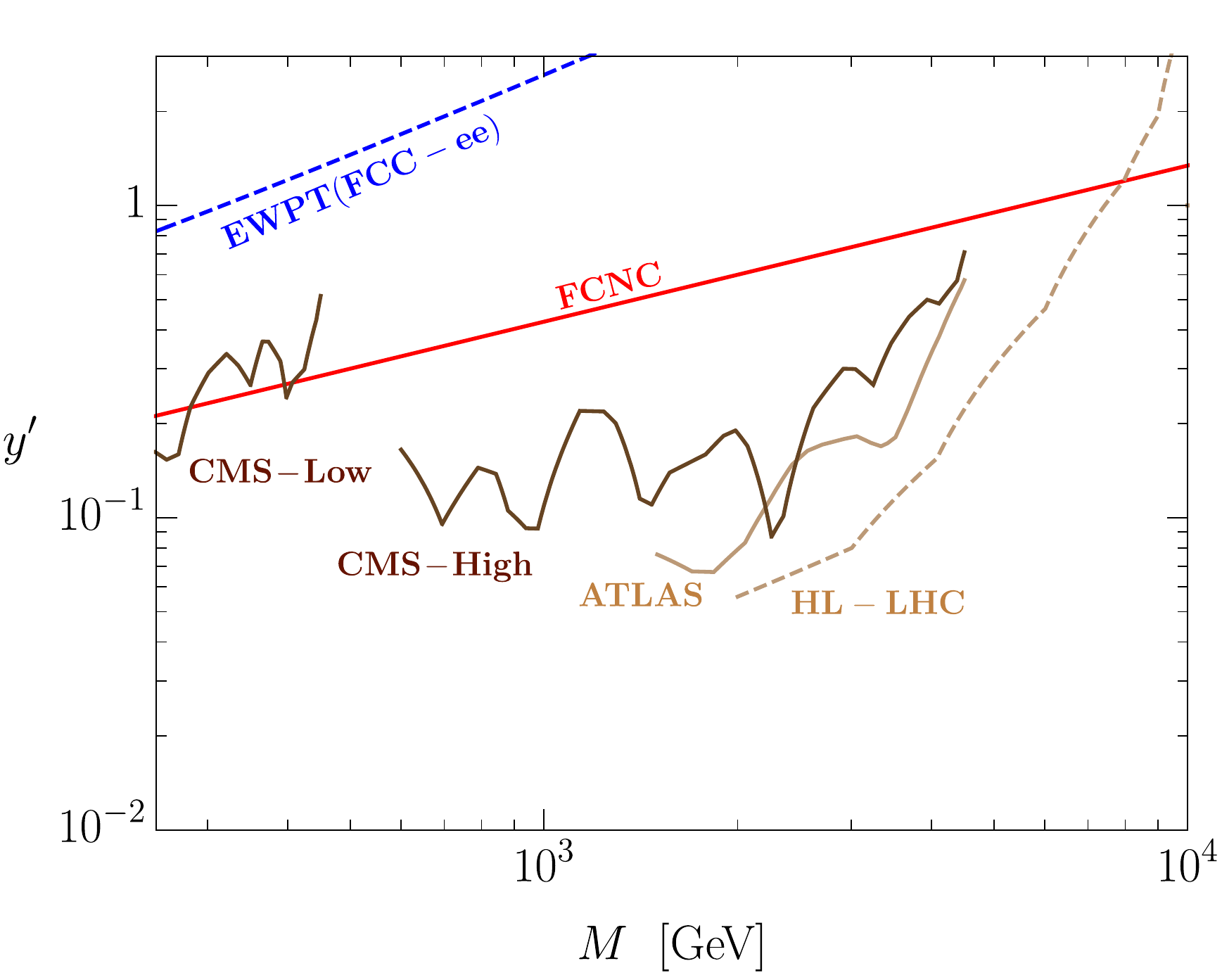}
\caption{\label{fig:scalarhighenplots}
Constraints on scalar doublet completion in the $M-y'$ plane. Shown are current bounds from neutral kaon mixing (red solid line) and dijet searches search at the LHC  (brown solid lines), including high mass dijet searches ("ATLAS" and "CMS-High")~\cite{Aad:2019hjw,CMS:2018mgb,Sirunyan:2019vgj} and a low mass boosted dijet search ("CMS-Low")~\cite{Sirunyan:2019vxa}. The expected future reach from precision measurements of $R_\ell$ at FCC-ee (blue dashed line) and high mass dijet searches at the HL-LHC~\cite{ATL-PHYS-PUB-2015-004} (brown dashed line) are also indicated. The trilinear scalar coupling $\kappa$ is chosen to saturate its naturalness condition, which is the minimum of either Eqs.~(\ref{eq:kappabound1}) and (\ref{eq:kappabound2}), while the physical singlet scalar mass is set $m_s = 1$ GeV.
}
\end{figure}
%%%%%%%%%%%%%%%%%%%

Here we summarize the experimental constraints and prospects in the scalar doublet completion of the light up-specfic scalar. As discussed earlier in Sec.~\ref{sec:VLQ-naturalness}, for a light scalar satisfying naturalness conditions (\ref{eq:kappabound1},\ref{eq:kappabound2}), we typically expect $\kappa \ll y'$. In this case, the strongest bounds on the UV completion are on  the coupling $y'$ and the scalar doublet mass $M$. These limits are compiled in Fig.~\ref{fig:scalarhighenplots}, where we show the constraints from FCNCs in the neutral kaon system and direct searches for dijet resonances at the LHC. We also display the projected reach of precision measurments of the $Z$ boson hadronic width at FCC-ee and high-mass dijet searches at the HL-LHC. 

\begin{figure}
\centering
\includegraphics[width=0.49\textwidth]{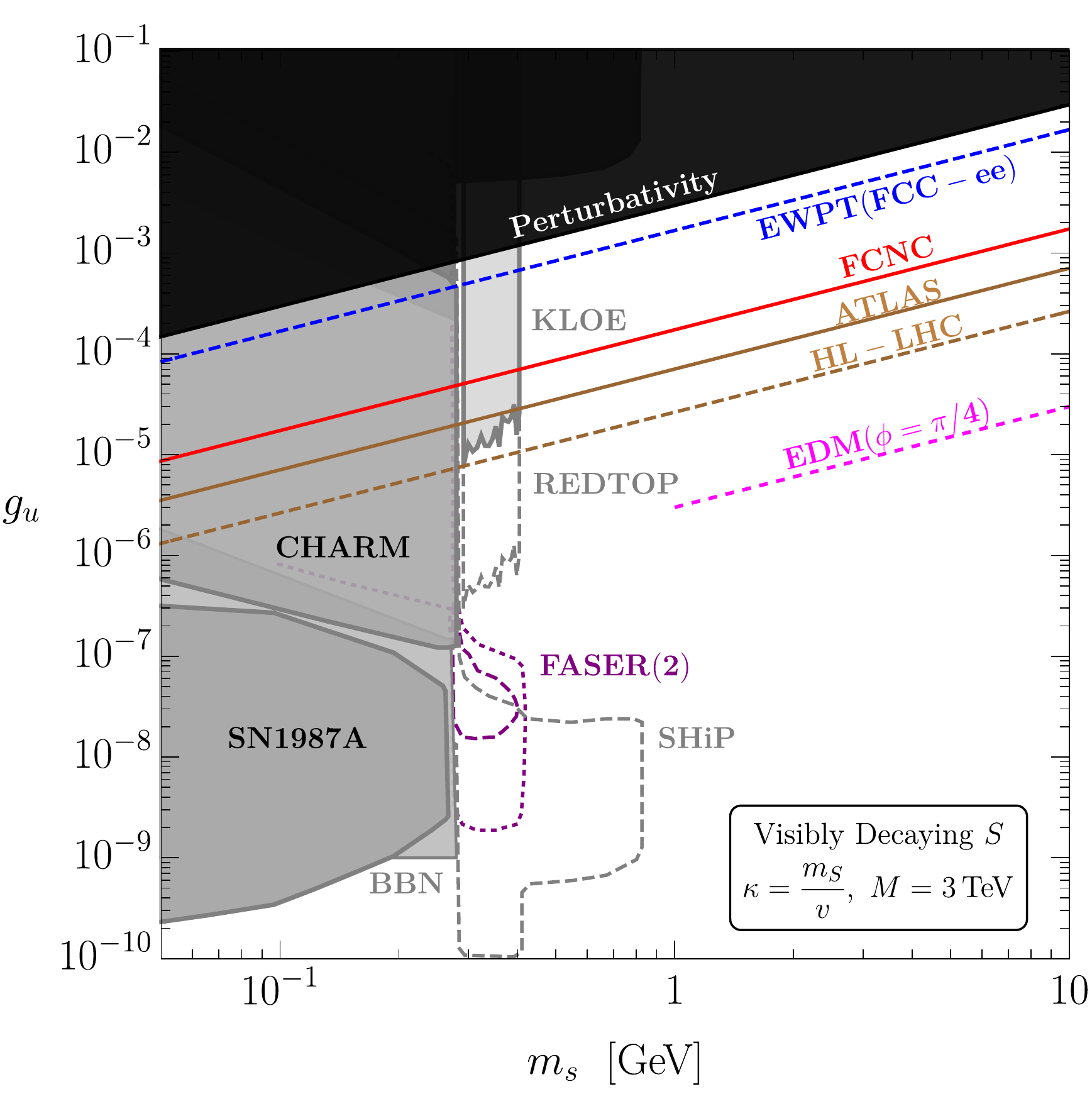}
\includegraphics[width=0.49\textwidth]{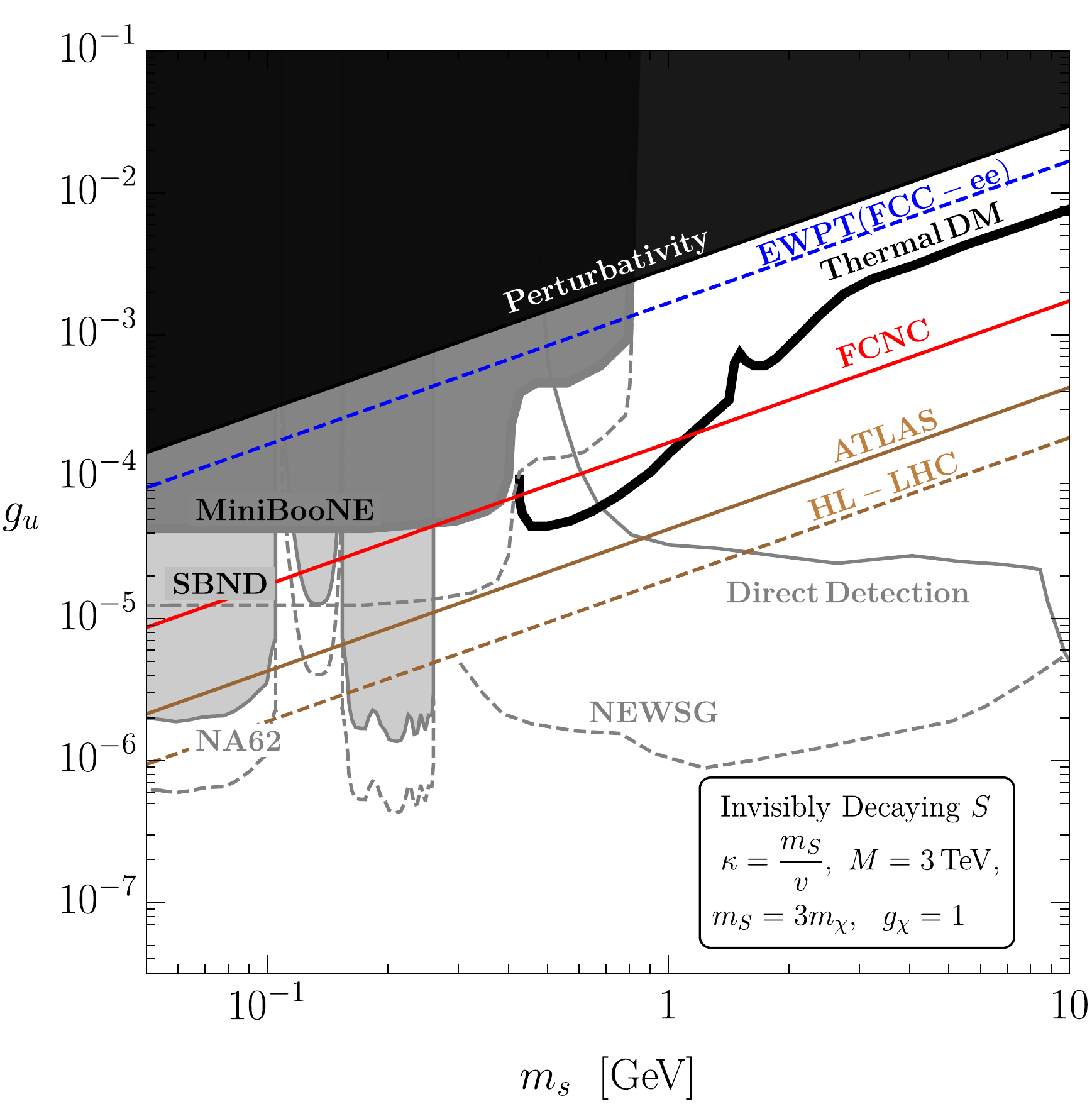}
\caption{\label{fig:scalarlowenplots}
 The up-specific scalar EFT parameter space shown in the $m_s-g_u$ plane. The left panel assumes the scalar decays visibly to hadrons, while the right panels assume the scalar decays invisibly to dark matter with $g_{\chi} = 1$ and $m_s = 3 m_\chi$. In both panels the coupling $y'$ is varied while the scalar doublet mass is fixed to $M = 3$ TeV and $\kappa$ is chosen to saturate the naturalness condition (\ref{eq:kappabound2}). In both panels we show several model-independent constraints from Ref.~\cite{Batell:2018fqo} on the EFT parameter space, which depend only on $g_u$ and $m_s$. In addition, constraints from the scalar doublet completion are shown under the stated assumptions for each plot. Further details are given in the main text.
}
\end{figure}

As was done for the VLQ completion, we interpret the bounds on the scalar doublet completion within the up-specific scalar EFT mass--coupling parameter space. Two interpretations are presented in Fig.~\ref{fig:scalarlowenplots}, where a number of bounds and projections are displayed in the $m_S$--$g_u$ plane. In particular, we show the model-independent constraints relying only on $g_u$ and $m_S$ derived previously in Ref.~\cite{Batell:2018fqo}; we refer the reader to Sec.~\ref{sec:VLQ-summary} for further details. Furthermore, we display the additional constraints that arise in the scalar doublet completion. The left panel assumes the scalar decays visibly to hadrons, while the right panel assumes the scalar decays invisibly to dark matter with $g_{\chi} = 1$ and $m_S = 3 m_\chi$. In both plots,  $y'$ is varied while the scalar doublet mass is fixed to $M = 3$ TeV and $\kappa$ is chosen to saturate the naturalness condition (\ref{eq:kappabound2}).  We see that the bounds from the scalar doublet completion cover interesting regions of the light scalar parameter space and as such complement those obtained by only considering up-specific EFT~\cite{Batell:2018fqo}. 

%%%%%%%%%%%%%%%%%%%%
%%%%%%%%%%%%%%

%%%%%%%%%%%%%%%%%%%%%%%%%%%%%%%%%%%%%%%%
%%%%%%%%%%%%%%%%%%%%%%%%%%%%%%%%%%%%%%%%
\section{Conclusions}
\label{sec:conclusions}

In this work we have studied two simple renormalizable completions of flavor-specific scalar mediators. While for concreteness we have focused on the up quark-specific coupling, similar models can straightforwardly be constructed for other flavor-specific couplings. In the first completion, a new VLQ mediates interactions between the light quarks, Higgs, and scalar singlet. In the second model, the interactions occur via a second scalar electroweak doublet. In both models we have studied the implications of naturalness on the size of the scalar potential and other couplings in the theory. A sizeable effective singlet--Higgs--quark coupling implies that the mediators (VLQ or scalar doublet) cannot be arbitrarily heavy, which opens new opportunities for experimental tests. We have derived bounds from the hadronic decay width of the $Z$ boson, FCNCs in the neutral kaon system, the neutron EDM, deviations in CKM unitarity, and direct searches for the new SM-charged states at the LHC. These models can be further tested at the HL-LHC and at future colliders. The bounds we derived can also be interpreted within the low energy flavor-specific EFT and are found to probe new regions in the scalar mass -- effective coupling plane. This underscores the general expectation that renormalizable completions can provide complementary constraints and new experimental opportunities to probe flavor-specific scalars. 

Looking ahead, there is significant scope for further phenomenological exploration within the flavor-specific framework. Investigations of other flavor-specifc couplings beyond the up quark-specific one studied here and in \cite{Batell:2018fqo} and the muon-specific one studied in \cite{Batell:2017kty} would be valuable and are likely to present new opportunities for model building (e.g., as a mediator to dark matter) and novel experimental prospects. In addition, it would be interesting to consider the UV origin of the flavor specific hypothesis, which may ultimately be tied to the dynamics underlying the SM flavor structure. 
%%%%%%%%%%%%%%%%%%%%%%%%%%%%%%%%%%%%%%%%
%%%%%%%%%%%%%%%%%%%%%%%%%%%%%%%%%%%%%%%%
\acknowledgments

We thank Jordy de Vries, Daniel Egana-Ugrinovic, Samuel Homiller, and Patrick Meade for helpful discussions and correspondence.  The work of B.B.~and M.R.~is supported by the U.S.~Department of Energy under grant No. DE–SC0007914. This work of A.F.~is supported in part by the National Science Foundation under Grant No.~PHY-1820760. The work of A.I.~is supported in part by the U.S.~Department of Energy under Grant No.~DE-SC0016013. The work of D.M.~is supported by the Natural Sciences and Engineering Research Council of Canada. TRIUMF receives federal funding via a contribution agreement with the National Research Council Canada.

%%%%%%%%%%%%%%%%%%%%%%%%%%%%%%%%%%%%%%%%
%%%%%%%%%%%%%%%%%%%%%%%%%%%%%%%%%%%%%%%%
\appendix
\numberwithin{equation}{section}

%%%%%%%%%%%%%%%%%%%%%%%%%%%%%%%%%%%%%%%%
%%%%%%%%%%%%%%%%%%%%%%%%%%%%%%%%%%%%%%%%
\section{Flavor-specific hypotheses in renormalizable completions}
\label{app:hypothesis}

%%%%%%%%%%%%%%%%%%%%%%%%%%%%%%%%%%%%%%%%
\subsection{VLQ model}
\label{app:VLQ-flavor-hypothesis}

The Lagrangian of the VLQ model is 
\begin{align}
\mathcal{L}& \supset i \overline{Q}_L \slashed{D} \, Q_L + i \overline{{\cal U}}_R \slashed{D} \, {\cal U}_R + i \overline{d}_R \slashed{D} \, d_R + i \overline{U}'_L \slashed{D} \, U'_L \nonumber \\
&- \left( \overline{Q}_L {\cal Y}_u\, {\cal U}_R H_c + \overline{Q}_L Y_d \, d_R H + \overline U'_L M\, {\cal U}_R  +\overline U'_L \lambda\, {\cal U}_R S + \mathrm{h.c.} \right),
\end{align}
where we have defined the fourplet ${\cal U}_R^\top \equiv (u_R, U'_R)$. In the limit of vanishing ${\cal Y}_u, Y_d, M, \lambda$, there is a large global flavor symmetry
\begin{equation}
    G = U(3)_Q \times U(4)_{{\cal U}_R} \times U(3)_{d_R}\times  U(1)_{U'_L}.
\end{equation}
The up-specific hypothesis can be understood by promoting the couplings ${\cal Y}_u, Y_d, M, \lambda$ to spurions and specifying how their background values explicitly break the symmetry $G$:
\begin{align}
{\cal Y}_u &\sim ({\bf 3},{\bf \overline 4},{\bf 1}, 0 ), ~~~~ G \rightarrow U(1)_{Q_1+u_R+U'_R} \times U(1)_c\times U(1)_t \times U(3)_{d_R} \times U(1)_{U'_L} ,\\
Y_d &\sim ({\bf 3},{\bf 1},{\bf \bar 3}, 0 ), 
~~~~ G \rightarrow U(1)_d \times U(1)_s\times U(1)_b \times U(4)_{{\cal U}_R} \times U(1)_{U'_L} ,\\
M &\sim ({\bf 1},{\bf \bar 4},{\bf 1}, 1 ), ~~~~ G \rightarrow U(3)_Q \times U(3)_{u_R}\times U(3)_{d_R} \times U(1)_{U'_L+U'_R}, \\
\lambda &\sim ({\bf 1},{\bf \bar 4},{\bf 1}, 1 ), ~~~~ G \rightarrow U(3)_Q \times U(3)_{d_R}\times U(3)_{c_R+t_R+U'_R} \times U(1)_{U'_L+u_R} .
\end{align}
With all spurions set to their background values, the full flavor symmetry is broken to a generalized baryon number under which all quark fields, including the VLQs, are charged. By performing suitable $G$ rotations we arrive at the starting Lagrangian in the main text, Eq.~(\ref{eq:L-VLQ-UR}).

%%%%%%%%%%%%%%%%%%%%%%%%%%%%%%%%%%%%%%%%
\subsection{Scalar doublet model}
\label{app:SD-flavor-hypothesis}

The Lagrangian of the scalar doublet model is given by
\be
\mathcal{L} \supset i \overline{Q}_L \slashed{D} Q_L + i \overline{{ u}}_R \slashed{D} { u}_R + i \overline{d}_R \slashed{D} d_R - \left( \overline{Q}_L Y_u { u}_R H_c + \overline{Q}_L Y_d d_R H + \overline{Q}_L Y'_u { u}_R H'_c
+\mathrm{h.c.} \right).
\ee
 In the limit of vanishing $Y_u, Y_d, Y'_u$, there is a large global flavor symmetry (the same as in the SM):
\begin{equation}
    G = U(3)_{Q_L}\times U(3)_{ u_R} \times U(3)_{d_R}.
\end{equation}
To define the up-specific hypothesis, we specify how the spurions $Y_u, Y_d, Y'_u$  explicitly break the symmetry $G$:
\begin{align}
Y_u &\sim ({\bf 3},{\bf \overline 3},{\bf 1}), ~~~~ G \rightarrow U(1)_u \times U(1)_c\times U(1)_t ,\\
Y_d &\sim ({\bf 3},{\bf 1},{\bf \bar 3}), 
~~~~ G \rightarrow U(1)_d \times U(1)_s\times U(1)_b, \\
Y'_u &\sim ({\bf 3},{\bf \overline 3},{\bf 1}), ~~~~ G \rightarrow U(1)_u \times U(2)_{ctL} \times U(2)_{ctR}, 
\end{align}
With all couplings assuming their background values, the only remaining global symmetry present in the theory is baryon number. In the main text, the coupling $y'$ in Eq.~(\ref{eq:L-HH}) is identified with the coupling  $Y'_u$ discussed here.

%%%%%%%%%%%%%%%%%%%%%%%%%%%%%%%%%%%%%%%%
%%%%%%%%%%%%%%%%%%%%%%%%%%%%%%%%%%%%%%%%
\section{VLQ with complex couplings}
\label{app:VLQ-scalar-couplings}

Here we consider general complex phases for the new physics couplings in the VLQ model. After transforming the quarks to the SM basis, there is a mass mixing described by the Lagrangian
\begin{align}
-\Lagr & =\left(\begin{array}{cc}
\overline {u}_{L} & \overline {U}'_{L}\end{array}\right)\left(\begin{array}{cc}
\displaystyle{\frac{y_u  v}{\sqrt{2}}} & \displaystyle{\frac{y v}{\sqrt{2}}}\\
\lambda v_S & M
\end{array}\right)\left(\begin{array}{c}
{u}_{R}\\
{U}'_{R}
\end{array}\right) +{\rm h.c.} \\
& = \overline \psi_L  \, {\cal M} \, \psi_R +{\rm h.c.} \nonumber
\end{align}
where in the second line we have defined $\psi_{L,R}^T  = (u_{L,R}, ~ U'_{L,R})$ and the mass matrix ${\cal M}$ in the obvious way. To diagonalize the system, we perform separate unitary transformations on the quark fields, 
\begin{equation}
\psi_L \rightarrow  L \, \psi_L, ~~~~  \psi_R \rightarrow R \, \psi_R, 
\end{equation}
where $L,R$ are unitary matrices satisfying ${\cal M}_u^D = L^\dag \, {\cal M}  \, R = {\rm diag}(m_u, m_{U'})$. 

We now consider the interactions. In the gauge sector, we obtain the following couplings involving the $W$ boson in the physical basis:
\begin{equation}
{\cal L} \supset \frac{g}{\sqrt{2}} W_\mu^+ \left(  L_{11}^* \, V_{1i}  \, \overline u_L \gamma^\mu d_{Li}  +  L_{12}^* \, V_{1i} \, \overline U'_L \gamma^\mu d_{Li}     \right)  + {\rm h.c.},
\end{equation}
where $i = 1,2,3$ runs over the three SM generations. The $Z$ boson couplings in the $\psi_L$ sector are 
\begin{align}
{\cal L} &  \supset \frac{g}{c_W} Z_\mu \bigg\{  \overline u_L \gamma^\mu \left( \tfrac{1}{2} L_{11}^* L_{11} -\tfrac{2}{3} s_W^2   \right)  u_{L}   +
\left[  \overline u_L \gamma^\mu \left( \tfrac{1}{2} L_{11}^* L_{12} \right)  U'_{L}  +{\rm h.c.} \right]  \\  \nonumber 
& ~~~ +   \overline U'_L \gamma^\mu \left( \tfrac{1}{2} L_{12}^* L_{12} -\tfrac{2}{3} s_W^2   \right)  U'_{L}    \bigg\} 
\end{align}
while those in the $\psi_R$ sector are unmodified. Note that without loss of generality, the phases in the elements $L_{11}, L_{12}$ can be removed by phase rotations of $u_{L,R}$ and $U_{L,R}'$, such that no new phases appear in the weak boson interactions.

Next, considering the scalar-fermion sector, we find 
\begin{equation}
-{\cal L} = Y_{h \bar \psi_I \psi_J} \, h \,  \overline \psi_{LI}  \, \psi_{RJ} + Y_{S \bar \psi_I \psi_J}  \, S \, \overline \psi_{LI}  \, \psi_{RJ}  +  {\rm h.c.},
\end{equation}
where we have defined the couplings 
\begin{align}
Y_{h \bar \psi_I \psi_J} & =   \frac{1}{\sqrt{2}} L_{1 I}^* \left( y_u R_{1J} + y R_{2J}.  \right)  \\
Y_{S \bar \psi_I \psi_J} & =  \lambda \, L_{2 I}^* \,R_{1J}.
\label{eq:S-quark-couplings}
\end{align}
Here $I, J = 1, 2$ for the light SM up quark and VLQ, respectively. 

We note that if any one of the couplings $M$, $y$, $\lambda$ or $y_u$ vanishes, the new complex phase is unphysical and can be removed through suitable rotations of the quark fields. 

\bibliography{flavor-specific}

\end{document}